\documentclass[12pt]{article}
\usepackage{times}
\usepackage[english]{babel}
\usepackage{bm}
\usepackage{multirow}
\usepackage{multicol}
\usepackage{url}
\usepackage{amsmath}
\usepackage{commath}
\usepackage{authblk}
\usepackage{txfonts}
\usepackage{xcolor, soul}
\sethlcolor{white}
\usepackage{times}
\usepackage{bm}
\usepackage{natbib}
\usepackage{tikz}
\usetikzlibrary{arrows,positioning,shapes.geometric,decorations.markings}
\tikzstyle{squarenode}=[rectangle,draw]
\tikzstyle{littlenode}=[circle,draw,minimum size=0.8cm,font=\small]
\tikzstyle{bigellipse}=[ellipse,draw,x radius=10cm, y radius=5cm]
\tikzset{negate/.style={
            decoration={markings,
            mark= at position 0.30 with {
                  \node[yshift=13pt,transform shape] (tempnode) {$\Bigg\Vert$};
                  }
              },
              postaction={decorate}
}
}
\usepackage{setspace}
\usepackage{subfig}
\usepackage{tabularx}
\usepackage{multirow}
\usepackage{pgfplots}
\usepackage{booktabs}
\usepackage[plain,noend]{algorithm2e}

\makeatletter
\renewcommand{\algocf@captiontext}[2]{#1\algocf@typo. \AlCapFnt{}#2} 
\def\@algocf@capt@plain{top}
\renewcommand{\algocf@makecaption}[2]{%
  \addtolength{\hsize}{\algomargin}%
  \sbox\@tempboxa{\algocf@captiontext{#1}{#2}}%
  \ifdim\wd\@tempboxa >\hsize
    \hskip .5\algomargin%
    \parbox[t]{\hsize}{\algocf@captiontext{#1}{#2}}
  \else%
    \global\@minipagefalse%
    \hbox to\hsize{\box\@tempboxa}
  \fi%
  \addtolength{\hsize}{-\algomargin}%
}
\makeatother

\definecolor{lightgray}{rgb}{1.0, 1.0, 1.0}

\def\ind{\perp\!\!\!\perp}

\newcommand{\cpr}{\mbox{cpr}}

\newcommand{\tu}{\textup}

\newcommand{\rhs}{\mbox{rhs}}

\newcolumntype{Y}{>{\centering\arraybackslash}X}

\begin{document}





\title{Path Analysis for Binary Random Variables}

\author[1,2]{Martina Raggi}
\author[3]{Elena Stanghellini}
\author[4]{Marco Doretti}

\affil[1]{University of Neuch\^{a}tel, Faculty of Economics and Business}
\affil[2]{Université de Paris, Inserm U1153, Epidemiology of Ageing and Neurodegenerative Diseases}
\affil[3]{University of Perugia, Department of Economics}
\affil[4]{University of Perugia, Department of Political Science}

\date{}

\maketitle

\vspace{-1.6cm}

\begin{center}
\footnotesize{\emph{A version of this paper is forthcoming in Sociological Methods \& Research.}}
\end{center}

\begin{abstract}
\footnotesize{The decomposition of the overall effect of a treatment into direct and indirect effects is here investigated with reference to a recursive system of binary random variables. We show how, for the single mediator context, the marginal effect measured on the log odds scale can be written as the sum of the indirect and direct effects plus a residual term that vanishes under some specific conditions. We then extend our definitions to situations involving multiple mediators and address research questions concerning the decomposition of the total effect when some mediators on the pathway from the treatment to the outcome are marginalized over. Connections to the counterfactual definitions of the effects are also made. Data coming from an encouragement design on students’ attitude to visit museums in Florence, Italy, are reanalyzed. The estimates of the defined quantities are reported together with their standard errors to compute p-values and form confidence intervals.}

\end{abstract}


\newpage

\section{Introduction}\label{sec:intro}
The decomposition of the total effect of a treatment $X$ on an outcome variable $Y$ into direct and indirect effects is a central topic in empirical research. In linear models, the relationship between total, direct and indirect effects is well understood (\Citealt{Cochran1938, alwin1975decomposition,baron1986moderator,bollen1987total}) and a simple decomposition is available. Such a decomposition is based on the linearity of the marginal model for $Y$ against $X$, where the coefficient of $X$ is equal to the sum of the direct effect and the indirect effects. Outside the linear case, this simplicity is lost, as in the marginal model of $Y$ against $X$ only, either the effect of $X$ on $Y$ is a complex function of the original parameters or the error term does not possess nice properties, or both. 

We here consider situations where the outcome $Y$ is a binary random variable. Contributions have addressed the case of one continuous mediator, see for example~\cite{mackinnon2007intermediate};~\Citet*{karlson2012comparing,breen2013total,breen2018note}.
Recent results concern the exact parametric form of the marginal effect of $X$ on $Y$ on the log odds ratio scale when the mediator is also binary~(\Citealt{stanghellini2019marginal}). In this setting, when $X$ is continuous the marginal model of $Y$ against $X$ is non-linear unless some conditional independence assumptions hold, and a rather complex formula links the marginal and conditional effect of $X$ on $Y$.  Similarly, for a discrete $X$ the parameters of the conditional model combine in a non-linear fashion to form the marginal effect. For analogous results on the log relative risk scale see~\cite{lupparelli2019conditional}.

Starting from the results in~\cite{stanghellini2019marginal}, we here elaborate a novel proposal for the direct and indirect effect definitions on the log odds scale for a treatment variable $X$ either continuous or discrete. The postulated system can be represented by a Directed Acyclic Graph (DAG), see~\Citet*[Ch. 2]{lauritzen1996graphical}, to which we refer for definitions; see also~\cite{elwert2013graphical} for an account in the sociological context. Our proposal is based on zeroing the path-specific regression coefficients. Graphically, this corresponds to deleting one arrow in the associated DAG and thereby represents the analogue of the path analysis method. 

We initially focus on a single mediator context and show that the marginal effect can be written as the sum of the indirect and direct effects plus a residual term that vanishes under some specific conditions. The proposed parametric relationship allows, for the specific setting under investigation, to solve the debate on which method should be used to disentangle the total effect, i.e., the product method or the difference method~(\Citealt{breen2018note}). It also avoids fitting two nested models, thereby sidestepping the issue of unequal variance~(\Citealt{winship1983structural}).
We then extend our derivations to the case of multiple binary mediators, also modeled as a recursive system of univariate logistic regressions. In this context, additional path-specific effects can be defined and different research questions addressed. Although the paper draws from the derivation in~\mbox{\cite{stanghellini2019marginal}}, some novel results are also presented. With reference to a single mediator, a general formulation of functional form linking the log odds ratio of the mediator and of the outcome to the covariates is considered. This is then extended to the multiple mediator context, for which a strategy for deriving the direct and indirect effects when marginalizing over an intermediate or outer mediator is also illustrated.

Our approach is developed in a purely associational context that, in general, holds no interpretation for causal inference. However, if the recursive system of equation is structural~(\Citealt[Ch. 7]{PearlCausalityBook2009}) and no unmeasured confounders exist, the total effect and some of its components can be endowed with a causal interpretation. Notice that a decomposition of the total effect based on counterfactual entities has been given by~\cite{Pearl2001,pearl2012mediation} and extended to the odds ratio scale by~\cite{VdWVan2010}. This parallelism is also addressed in this paper.

In Section~\ref{sec:singlew}, we offer the general theory for the case of a single mediator. A case study concerning a randomized encouragement experiment on cultural consumption performed in Florence (Italy) is also presented as a guiding example. Results of a simulation study investigating maximum likelihood (ML) estimation of the effects and other related measures is also reported in Section~\mbox{\ref{sec:sims}}, while the extension to the multiple mediator setting is contained in Section~\ref{sec:multiplew}. In Section~\ref{sec:other}, we address other complex issues concerning path-specific effects, whereas links with counterfactual definitions are explored in Section~\ref{sec:causal}. Finally, in Section~\ref{sec:concl} we draw some conclusions. 


\begin{figure}[tb]
\centering{
\subfloat[][]{\label{fig:dag1}
\begin{tikzpicture}[scale=0.4,auto,->,>=stealth',shorten >=1pt,node distance=2cm] 
\node[littlenode] (W) {$W$};
\node[littlenode] (X) [below left of=W] {$X$};
\node[littlenode] (Y) [below right of=W] {$Y$};
\draw[->] (W) --node {} (Y) ; \draw[->] (X) --node {} (W); \draw[->] (X) --node {} (Y);
\end{tikzpicture}}\quad
\subfloat[][]{\label{fig:dag2}
\begin{tikzpicture}[scale=0.4,auto,->,>=stealth',shorten >=1pt,node distance=2cm] 
\node[littlenode] (W) {$W$};
\node[littlenode] (X) [below left of=W] {$X$};
\node[littlenode] (Y) [below right of=W] {$Y$};
\draw[->] (X) --node {} (W); \draw[->] (W) --node {} (Y);
\end{tikzpicture}}\\
\subfloat[][]{\label{fig:dag3}
\begin{tikzpicture}[scale=0.4,auto,->,>=stealth',shorten >=1pt,node distance=2cm] 
\node[littlenode] (W) {$W$};
\node[littlenode] (X) [below left of=W] {$X$};
\node[littlenode] (Y) [below right of=W] {$Y$};
\draw[->] (X) --node {} (W); \draw[->] (X) --node {} (Y);
\end{tikzpicture}}\quad
\subfloat[][]{\label{fig:dag4}
\begin{tikzpicture}[scale=0.4,auto,->,>=stealth',shorten >=1pt,node distance=2cm] 
\node[littlenode] (W) {$W$};
\node[littlenode] (X) [below left of=W] {$X$};
\node[littlenode] (Y) [below right of=W] {$Y$};
\draw[->] (W) --node {} (Y); \draw[->] (X) --node {} (Y);
\end{tikzpicture}}
}
\caption{Data generating process when (a) no conditional independences hold, (b) $X\ind Y\mid W$, (c) $W\ind Y\mid X$  and (d) $W\ind X$.\label{fig:med}}
\end{figure}
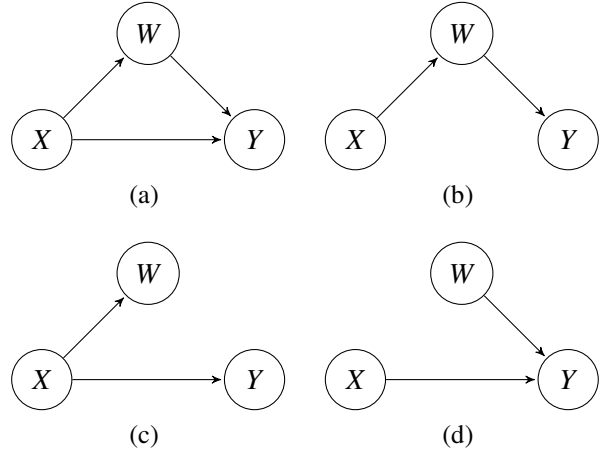

\section{Effect decomposition with a single mediator}\label{sec:singlew}
We first focus on a very simple model for a binary outcome $Y$, a binary mediator $W$ and a treatment $X$, that can be either discrete or continuous; see Fig~\ref{fig:med}(a) for the corresponding DAG. Our aim is to decompose the total effect of $X$ on $Y$ on the log odds scale. Our postulated models are a logistic regression for $Y$ given $X$ and $W$ and for $W$ given $X$, that is,
\begin{equation}\label{eq:lry}
\log\frac{P(Y=1\mid X=x,W=w)}{P(Y=0\mid X=x,W=w)} = \beta_{0}+\beta_{x}x+\beta_{w}w+\beta_{xw}xw
\end{equation}
and
\begin{equation}\label{eq:lrw}
\log\frac{P(W=1\mid X=x)}{P(W=0\mid X=x)} = \gamma_{0}+\gamma_{x}x.
\end{equation}
Notice that we allow for the interaction between $X$ and $W$ in the outcome equation. In order to make the paper self-contained, the derivations in~\cite{stanghellini2019marginal} to evaluate the total effect as a function of the parameters of models \eqref{eq:lry} and \eqref{eq:lrw} are here reproduced, prior the introduction of its decomposition into direct, indirect and residual effects. 

As a guiding example through this section, we consider an experiment aiming at identifying the best incentives to offer high school students in Florence to enhance cultural interest and increase art museum attendance. Three treatment levels are considered: a flyer given to the students  with the main information about the Palazzo Vecchio museum constitutes the first level; a flyer and a presentation of the museum from an expert constitutes the second level; a flyer, the presentation and a reward in the form of extra-credit points for their final school grade constitutes the third level. All students receive a free entry ticket to Palazzo Vecchio. The aim of the experiment is not only to assess the total effect of the treatment ($X$) on students museum's attendance ($Y$), but also to understand to what extent this effect could be stimulated by student's visit to Palazzo Vecchio ($W$); see~\mbox{\Citet*{lattarulo2017nudging}} and~\mbox{\cite{forastiere2019exploring}}.

The interest is in the marginal model of $Y$ against $X$, as a function of the parameters in \mbox{\eqref{eq:lry}} and \mbox{\eqref{eq:lrw}}. From first principles of probabilities, it follows that:
\begin{equation}\label{eq:wkfy}
\log\frac{P(Y=1\mid X=x)}{P(Y=0\mid X=x)} = -\log\frac{P(W=w\mid Y=1,X=x)}{P(W=w\mid Y=0,X=x)} + \log\frac{P(Y=1\mid W=w,X=x)}{P(Y=0\mid W=w,X=x)}.
\end{equation}
The second term of the right hand side (rhs) of the above equality is given from model \mbox{\eqref{eq:lry}}, while the parametric expression of the first term is not immediately derived from models \mbox{\eqref{eq:lry}} and \mbox{\eqref{eq:lrw}}, as it involves the probability of $W$ after conditioning on $X$ and $Y$ (not after conditioning on $X$ only). 
However, by repeated use of the previous relationship, we have:
\begin{equation}\label{eq:wkfw}
\log\frac{P(W=1\mid Y=y,X=x)}{P(W=0\mid Y=y,X=x)} = \log\frac{P(Y=y\mid W=1,X=x)}{P(Y=y\mid W=0,X=x)} + \log\frac{P(W=1\mid X=x)}{P(W=0\mid X=x)}.
\nonumber
\end{equation}
Using \mbox{\eqref{eq:lry}} and \mbox{\eqref{eq:lrw}}, after some simplifications, we find:
\begin{equation}\label{eq:gyx1}
\log\frac{P(W=1\mid Y=y, X=x)}{P(W=0\mid Y=y, X=x)} \\
= y(\beta_{w}+\beta_{xw}x) +\log\frac{1+\exp (\beta_{0}+\beta_{x}x)}{1+\exp (\beta_{0}+\beta_{x}x+\beta_{w}+\beta_{xw}x)} + \gamma_{0}+\gamma_{x}x.
\end{equation}
In what follows we denote with $g_{y}(x)$ the $\log\frac{P(W=1\mid Y=y, X=x)}{P(W=0\mid Y=y, X=x)}$, i.e.
\begin{equation}\label{eq:gyx}
\begin{split}
g_{y}(x) &= \log\frac{P(W=1\mid Y=y, X=x)}{P(W=0\mid Y=y, X=x)} \\
	     &= y(\beta_{w}+\beta_{xw}x) +\log\frac{1+\exp (\beta_{0}+\beta_{x}x)}{1+\exp (\beta_{0}+\beta_{x}x+\beta_{w}+\beta_{xw}x)} + \gamma_{0}+\gamma_{x}x.
\end{split}
\end{equation}
  
\noindent Since $(1+\exp g_{y}(x))^{-1}$ corresponds to $P(W=0 \mid Y=y, X=x)$, substituting in \mbox{\eqref{eq:wkfy}} for $w=0$, we find:

\begin{equation}\label{eq:wkfymar}
\log\frac{P(Y=1\mid X=x)}{P(Y=0\mid X=x)} =  \log\frac{1+\exp g_{1}(x) }{1+\exp g_{0}(x) } + \beta_{0}+\beta_{x}x.
\end{equation}

\noindent Denoting with
\[
\tu{\tu{RR}}_{W\mid Y, X=x}=\frac{P(W=1\mid Y=1,X=x)}{P(W=1\mid Y=0,X=x)}
\]
the relative risk of $W$ for varying $Y$ in the distribution of $X=x$, we have
\[
\tu{\tu{RR}}_{W\mid Y, X=x}=\frac{\exp g_1(x) \{1+\exp g_{0}(x)\} }{\exp g_0(x) \{1+\exp g_{1}(x)\} }.
\]
Analogously, letting $\bar W=1-W$, then 
\[
\tu{\tu{RR}}_{\bar W\mid Y, X=x}=\frac{1+\exp g_{0}(x) }{1+\exp g_{1}(x) }.
\]
It then follows that~\eqref{eq:wkfymar} can be rewritten as
\begin{equation}\label{eq:wkfymar1}
\log\frac{P(Y=1\mid X=x)}{P(Y=0\mid X=x)} =  \beta_{0}+\beta_{x}x-\log \tu{\tu{RR}}_{\bar W \mid Y, X=x}
\end{equation}
or, alternatively, as
\begin{equation}\label{eq:wkfymar2}
\log\frac{P(Y=1\mid X=x)}{P(Y=0\mid X=x)} = \beta_{0}+\beta_{x}x+ \beta_w+\beta_{xw}x -\log \tu{\tu{RR}}_{W \mid Y, X=x}.
\end{equation}

Notice that conditioning on a set of covariates $C=(C_1, \ldots C_p)$ does not strongly alter the structure of~\eqref{eq:wkfymar} and~\eqref{eq:wkfymar2}. We here offer an example for $p=1$, with both additive and interaction effects up to the second order in both models for $Y$ and $W$. After the marginalization over $W$, we obtain the marginal model for $Y$ given $X$ and $C$ as:

\begin{equation}\label{eq:wkfymarC}
\log\frac{P(Y=1\mid X=x,C=c)}{P(Y=0\mid X=x,C=c)} =  \beta_{0}+\beta_{x}x + \beta_c c + \beta_{xc} xc -\log \tu{RR}_{\bar W \mid Y, X=x, C=c}
\end{equation}
with 
\[
\tu{RR}_{\bar W\mid Y, X=x,C=c}=\frac{1+\exp g_{0}(x,c) }{1+\exp g_{1}(x,c)},
\]
and
\begin{equation}
\begin{aligned} \label{gfordati}
g_{y}(x,c)&= y \left( \beta_w + \beta_{xw} x + \beta_{cw} c \right) \\
&+ \log \frac{1+ \exp (\beta_0 + \beta_{x} x + \beta_{c} c + \beta_{xc}xc)}{1 + \exp (\beta_0 + \beta_x x + \beta_ c c + \beta_w + \beta_{xw} x + \beta_{xc} xc + \beta_{cw} c)}\\
& + \gamma_0 + \gamma_x x + \gamma_c c + \gamma_{xc} xc.
\end{aligned}
\end{equation}
See Appendix~\ref{app:cov} for a general formulation that includes more covariates and possibly nonlinear link functions.

\begin{table}[h]
\centering
\begin{tabular}{lllllllll}
 & \multicolumn{1}{l|}{} & $Y$ &  &  &  & \multicolumn{1}{l|}{} & $Y$ &  \\
$W=0$ & \multicolumn{1}{l|}{$C=0$} & 0 & 1 &  & $W=0$ & \multicolumn{1}{l|}{$C=1$} & 0 & 1 \\ \cline{1-4} \cline{6-9} 
$X$ & \multicolumn{1}{l|}{1} & 19 & 2 &  & $X$ & \multicolumn{1}{l|}{1} & 48 & 17 \\
 & \multicolumn{1}{l|}{2} & 14 & 21 &  &  & \multicolumn{1}{l|}{2} & 14 & 28 \\
 & \multicolumn{1}{l|}{3} & 3 & 3 &  &  & \multicolumn{1}{l|}{3} & 23 & 21 \\
 &  &  &  &  &  &  &  &  \\
 & \multicolumn{1}{l|}{} & Y &  &  &  & \multicolumn{1}{l|}{} & $Y$ &  \\
$W=1$ & \multicolumn{1}{l|}{$C=0$} & 0 & 1 &  & $W=1$ & \multicolumn{1}{l|}{$C=1$} & 0 & 1 \\ \cline{1-4} \cline{6-9} 
$X$ & \multicolumn{1}{l|}{1} & 0 & 0 &  & $X$ & \multicolumn{1}{l|}{1} & 1 & 2 \\
 & \multicolumn{1}{l|}{2} & 1 & 0 &  &  & \multicolumn{1}{l|}{2} & 6 & 3 \\
 & \multicolumn{1}{l|}{3} & 1 & 9 &  &  & \multicolumn{1}{l|}{3} & 19 & 11
\end{tabular}
\caption{Contingency tables for ($Y,X,W,C$) for the cultural consumption experiment.}
\label{tab:cont}
\end{table}

With reference to the cultural consumption data, 15 classes for a total of 294 students, all aged between 15 and 18, from three different schools,  were randomly assigned at baseline (March/April 2014) to the three treatment levels ($X$).  At the second occasion, after two months, researchers collected the entry tickets to record student visits ($W$). Finally, after six months they collected the concluding questionnaire with general information on visits to other museums ($Y$). A questionnaire with information on background characteristics of the students and their families was also administered. Among all the covariates, only one appears to be relevant in the model for the outcome, that is the binary variable $C$ taking value 1 for students considering themselves mainly interested in mathematics/science and 0 if they are mainly interested in humanities. At follow-up 28 students were absent, so the final sample included 266 students. Data are reported in Table~\mbox{\ref{tab:cont}} and are publicly available at~\url{https://www.tandfonline.com/doi/abs/10.1080/07350015.2019.1647843} as supplementary material of~\mbox{\cite{forastiere2019exploring}}.

Table~\mbox{\ref{tab:regr}} contains the output of the ML estimation of the logistic regression models for the outcome and for the mediator. We use the subscript $\{2,1\}$ to denote the contrast of level 2 (Flyer + Presentation) versus level 1 (Flyer) and $\{3,1\}$ for the contrast of level 3  (Flyer + Presentation + Reward) versus level 1. Notice that the interaction terms $\beta_{x_{\{2,1\}}w}$ and $\beta_{x_{\{3,1\}}w}$ in the outcome equation are significant.

\begin{table}[tb]
\centering
\begin{tabular}{lccccc}
\toprule
  \multicolumn{6}{c}{$Y\sim\beta_{0}+\beta_{x}X+\beta_{c}C+\beta_{w}W+\beta_{cw}CW$}   \\
 \cmidrule(l{3pt}r{3pt}){1-6}
 		& Estimate 	& Std. Error & \multicolumn{2}{c}{95\% Conf. Interval} 	& p-value \\
 \cmidrule(l{3pt}r{3pt}){2-6}
$\beta_{0}$             &  -1.6186  &   0.3857 &  -2.3746 &-0.8626 & 0.0000\\ 
$\beta_{x_{\{2,1\}}}$   &   1.9345  &   0.3676 &   1.2139 & 2.6550 & 0.0000\\ 
$\beta_{x_{\{3,1\}}}$   &   1.1329  &   0.3865 &   0.3754 & 1.8904 & 0.0034\\
$\beta_{c}$	            &   0.4597  &   0.3540 &  -0.2342 & 1.1536 & 0.1941\\  
$\beta_{w}$	            &   4.3290  &   1.5427 &   1.3053 & 7.3527 & 0.0050\\ 
$\beta_{x_{\{2,1\}}w}$  &  -3.7077  &   1.4725 &  -6.5937 &-0.8217 & 0.0118\\ 
$\beta_{x_{\{3,1\}}w}$  &  -2.2708  &   1.3365 &  -4.8903 & 0.3488 & 0.0893\\
$\beta_{cw}$            &  -2.4770  &   0.9255 &  -4.2910 &-0.6630 & 0.0074\\   
\midrule
 \multicolumn{6}{c}{$W\sim \gamma_{0}+\gamma_{x}X$} \\
 \cmidrule(l{3pt}r{3pt}){1-6}
		& Estimate 	& Std. Error & \multicolumn{2}{c}{95\% Conf. Interval} 	& p-value\\ 
 \cmidrule(l{3pt}r{3pt}){2-6} 
$\gamma_{0}$    &  -3.3557   &  0.5873    &  -4.5069 & -2.2046  & 0.0000 \\
$\gamma_{x_{\{2,1\}}}$	&   1.3145  &   0.6767     &  -0.0118 & 2.6409  & 0.0521  \\
$\gamma_{x_{\{3,1\}}}$	&   3.1326    &  0.6245       &    1.9087  & 4.3565  &   0.0000  \\
\bottomrule
\end{tabular}
\caption{ML estimates of the two logistic models for $Y$ and $W$ for the cultural consumption experiment.}
\label{tab:regr}
\end{table}

Ignoring for now the  sampling errors, we see that the $g_y(x,c)$ function can be formed by plugging in~\mbox{\eqref{gfordati}} the ML estimates of the parameters. The function expresses the log odds ratio of $W$ after conditioning on $X$, $Y$ and the covariate $C$.
\vspace{0.2cm}

\noindent 
We now present a definition for the total, direct and indirect effects in the situation with no covariates. When covariates $C$ are present, the parametric formula of $\tu{TE}(x)$ and its decomposition, for both continuous and discrete $X$, vary with the level of $C$. Notice that the direct and indirect effects so defined do not sum to the total effect, but a residual term remains. This term is zero only under some specific conditions, that we are going to discuss.
\vspace{0.2cm}

\noindent\textbf{Total Effect} 
\noindent Let $\tu{TE}(x)$ be the effect of $X$ on $Y$, on the log odds scale, in the distribution of $(Y\mid X)$ obtained after marginalization on $W$. For $X$ continuous and differentiable, the total effect is defined as the derivative of~\eqref{eq:wkfymar} with respect to $x$. For $X$ discrete, the total effect is defined as the difference between~ \eqref{eq:wkfymar} evaluated at two different levels of $X$. 

\vspace{0.2cm}
\noindent\textbf{Indirect Effect} 
\noindent Let $\tu{IE}(x)$ be the indirect effect of $X$ on $Y$ on the log odds scale. The indirect effect is defined as the part of the total effect of $X$ on $Y$ through $W$ only. It is evaluated after imposing, in the total effect, the coefficients of $X$ in the model for $Y$ equal to zero, i.e. $\beta_x=\beta_{xw}=0$, i.e. $\tu{IE}(x)=\tu{TE}(x)\mid_{\beta_x=\beta_{xw}=0}$, so that $X\ind Y\mid W$ and the effect of $X$ on $Y$ is mediated by $W$ (see Fig.~\ref{fig:med}(b)).

\vspace{0.2cm}
\noindent\textbf{Direct Effect}
\noindent Let $\tu{DE}(x)$ be the direct effect of $X$ on $Y$ on the log odds scale. The direct effect is defined as the part of the total effect due to $X$ only. It is obtained after imposing, in the total effect, the coefficients of $W$ in the model for $Y$ equal to zero, i.e. $\beta_w=\beta_{xw}=0$ so that $W\ind Y\mid X$ (see Fig.~\ref{fig:med}(c)). In other words, we have $\tu{DE}(x)=\tu{TE}(x)\mid_{\beta_w=\beta_{xw}=0}$. This definition is aligned with the collapsibility of odds ratio, as explained by \Citet*{xie2008some}, Corollary 3. It is important to notice that the direct effect can also be seen as the effect of $X$ on $Y$ keeping $W=0$.


\vspace{0.2cm}
\noindent\textbf{Residual Effect} 
\noindent  Let $\tu{RES}(x)$ be the residual effect of $X$ on $Y$ on the log odds scale defined as $\tu{TE}(x)-\tu{DE}(x)-\tu{IE}(x)$. Clearly, by construction
\begin{equation}\label{eq:tesum}
\tu{TE}(x)=\tu{DE}(x) + \tu{IE}(x) + \tu{RES}(x).
\end{equation}
Notice that this residual term is always null in linear models. In this context, the total effect can be decomposed into the sum of the direct and indirect effect. Provided that these two are positive, it is therefore meaningful to look at the ratio between the indirect effect and the total effect, as it gives an indication of the proportion of total effect due to the mediator $W$ (i.e., the proportion mediated). When a residual effect is present, the ratio between the indirect and total effect can still provide information on the weight of the indirect effect on the total effect, though with a less clear interpretation (see Sections~\mbox{\ref{subsec:singlewcont}} and~\mbox{\ref{subsec:singlewdisc}}).

\vspace{0.4cm} In what follows, we study in detail the decomposition of the total effect for the simple case without covariates, where $X$ can be either continuous or discrete. Addition of covariates can be done in a straightforward manner.

\subsection{Continuous case}\label{subsec:singlewcont}
\noindent We first look at the case of $X$ continuous and differentiable. Let 
$$\tu{TE}(x) = \frac{d}{d x} \log\frac{P(Y=1\mid X=x)}{P(Y=0\mid X=x)}.$$ 
It is possible to show that
\begin{equation}\label{eq:cochranint}
\begin{split}
\tu{TE}(x) & =  \beta_{x}\{1-\Delta_{y}(x)\Delta_{w}(x)\}\\
&+\beta_{xw}\{P(W=1|Y=1,X=x)-\Delta_{w}(x)P(Y=1\mid W=1,X=x)\}\\
&+\gamma_{x}\Delta_{w}(x)
\end{split}
\end{equation}
where
\[
\begin{split}
\Delta_{y}(x) &= P(Y=1\mid W=1,X=x)-P(Y=1\mid W=0,X=x) \\
&=\frac{\exp(\beta_0+\beta_x x + \beta_w + \beta_{xw} x)}{1+\exp(\beta_0+\beta_x x + \beta_w + \beta_{xw} x)} - \frac{\exp(\beta_0+\beta_x x)}{1+\exp(\beta_0+\beta_x x )}
\end{split}
\]
and
\[
\begin{split}
\Delta_{w}(x) &= P(W=1\mid Y=1,X=x)-P(W=1\mid Y=0,X=x) \\
&=\frac{\exp g_1(x)}{1+\exp g_1(x)}-\frac{\exp g_0(x)}{1+\exp g_0 (x)}
\end{split}
\]
with $g_y(x)$ as in~\eqref{eq:gyx}. Eq.~\eqref{eq:cochranint} confirms the well-known fact that the marginal logistic model is non linear in $x$, also providing the explicit expression of it. Notice that, as shown in~\cite{stanghellini2019marginal}, all terms in curly bracket are bounded between 0 and 1, while $\Delta_{w}(x)$ is bounded between -1 and 1. Notice further that $\Delta_w(x)$ and $\Delta_y(x)$ share the same sign, and they are both zero whenever $ W \ind  Y\mid X$.

\vspace{0.2cm}
\noindent {\bf Indirect Effect}
\noindent Following the definition, we evaluate the total effect assuming  ${\beta_x=\beta_{xw}=0}$, that is
\begin{equation}\label{eq:iecon}
\tu{IE}(x)=\tu{TE}(x)\mid_{\beta_x=\beta_{xw}=0}=\gamma_{x} \Delta_{w}^*(x)
\end{equation}
where $\Delta_{w}^*(x)$ is $\Delta_{w}(x)$ evaluated at $\beta_{x}=\beta_{xw}=0$. The indirect effect of $X$ on $Y$ through $W$ depends on the value of $x$, and is null if either $\gamma_x$ or $\beta_w$ are zero. It can be shown that, for all $x$, $\Delta_{w}^*(x)$ and $\beta_w$ share the same sign and, therefore, the indirect effect is concordant with the $\gamma_x\beta_w$ product; see \mbox{Appendix~\ref{app:delta}}. However, the magnitude of the effect varies with $x$.

\vspace{0.2cm}
\noindent {\bf Direct Effect} 
\noindent Following the definition, we evaluate the direct effect after assuming $\beta_w=\beta_{xw}=0$, that is
\begin{equation}\label{eq:decon}
\begin{split}
\tu{DE}(x)&=\tu{TE}(x)\mid_{\beta_w=\beta_{xw}=0}=\beta_{x}.\\
\end{split}
\end{equation}
Notice that~\eqref{eq:decon} follows as  $\Delta_y(x)$ and $\Delta_w(x)$ are zero when $\beta_w=\beta_{xw}=0$. 
 
\vspace{0.2cm}
\noindent {\bf Residual Effect} 
\noindent Finally, the residual effect is given by difference as follows
\begin{equation}\label{eq:ncecon}
\begin{split}
\tu{RES}(x)&=\tu{TE}(x)-\tu{IE}(x)-\tu{DE}(x)\\
& = - \beta_{x}\{\Delta_{y}(x)\Delta_{w}(x)\}\\
&+\beta_{xw}\{P(W=1|Y=1,X=x)-\Delta_{w}(x)P(Y=1\mid W=1,X=x)\}\\
&+\gamma_{x}\{\Delta_{w}(x)-\Delta^*_{w}(x)\}.
\end{split}
\end{equation}
It is therefore apparent that the effect above vanishes whenever $\beta_x=\beta_{xw}=0$ or $\beta_w=\beta_{xw}=0$. As a matter of fact, in the former case we have $\Delta_{w}(x)=\Delta^*_{w}(x)$, whereas in the latter case $\Delta_{w}(x)=\Delta_{y}(x)=0$. Notice that the latter case coincides with the condition of collapsibility of odds ratio, see \cite{xie2008some}, Corollary 3. Since all terms in curly brackets are bounded, expression ~\mbox{\eqref{eq:ncecon}} also highlights that the sign of the residual effect depends on the relative magnitude of the coefficients and is not linked to the logistic regression coefficients in a clear way. As a matter of fact, even if the direct and indirect effects share the same sign, the sign of the residual effect may be either positive or negative. Thus, as mentioned in Section~\mbox{\ref{sec:singlew}}, for a fixed level of $x$, the ratio between the indirect effect and the total effect may provide an indication of the relative strength of the indirect effect, though with a less clear interpretation.

\vspace{0.2cm}
\noindent \textbf{Some cases of interest}
\noindent Reformulating the total effect by definition as in \eqref{eq:tesum}, we study in detail the decomposition of the total effect into indirect \eqref{eq:iecon}, direct \eqref{eq:decon} and residual  \eqref{eq:ncecon} effects for some cases of interest.

\vspace{0.2cm}
\noindent \textbf{Case \textit{i})} When the recursive logistic models can be depicted as in Fig.~\ref{fig:med}(b), i.e. $X\ind Y\mid W$, it follows from the definition above that 
\[
\tu{TE}(x)\mid_{\beta_x=\beta_{xw}=0}=\tu{IE}(x)
\]
and both $\tu{DE}(x)$ and $\tu{RES}(x)$ are zero.
\vspace{0.2cm}

\noindent \textbf{Case \textit{ii})} When the recursive logistic models can be depicted as in Fig.~\ref{fig:med}(c), i.e. $W\ind Y\mid X$, it follows from the definition above that 
\[
\tu{TE}(x)\mid_{\beta_w=\beta_{xw}=0} = \tu{DE}(x)
\]
and both $\tu{IE}(x)$ and $\tu{RES}(x)$ are zero.

\vspace{0.2cm}
\noindent \textbf{Case \textit{iii})} A noticeable situation arises after imposing $\beta_{xw}=0$. In this case, the total effect is
\[
\tu{TE}(x)\mid_{\beta_{xw}=0}=\tu{DE}(x)+\tu{IE}(x)+\tu{RES}(x)\mid_{\beta_{xw}=0}
\]
where
\[
\tu{RES}(x)\mid_{\beta_{xw}=0}= - \beta_{x}\{\Delta_{y}(x)\Delta_{w}(x)\}\mid_{\beta_{xw}=0}+\gamma_{x}\{\Delta_{w}(x)-\Delta^*_{w}(x)\}\mid_{\beta_{xw}=0}.
\]
Notice that this assumption does not itself reflect into any conditional independence. If we further assume $\gamma_{x}=0$ then some simplifications arise. Thus
\[
\tu{TE}(x)\mid_{\beta_{xw}=\gamma_x=0}=\tu{DE}(x)+\tu{RES}(x)\mid_{\beta_{xw}=\gamma_{x}=0}
\]
where
\[
\tu{RES}(x)\mid_{\beta_{xw}=\gamma_x=0}= - \beta_{x}\{\Delta_{y}(x)\Delta_{w}(x)\}\mid_{\beta_{xw}=\gamma_x=0}.
\]
It is possible to see that under this condition $|\tu{TE}(x)| \leq |\beta_{x}|$, in line with results obtained by~\cite{neuhaus1993geometric} in a more general context.

\vspace{0.2cm}
\noindent \textbf{Case \textit{iv})} When the recursive logistic models can be depicted as in Fig.~\ref{fig:med}(d), i.e. $W\ind X$, it follows from the definition above that
\[
\tu{TE}(x)\mid_{\gamma_x=0} = \tu{DE}(x)+\tu{RES}(x)\mid_{\gamma_x=0}
\]
where
\begin{equation}\notag
\begin{split}
\tu{RES}(x)\mid_{\gamma_x=0}&= - \beta_{x}\Delta_{y}(x)\{\Delta_{w}(x)\}\mid_{\gamma_x=0}\\
&+\beta_{xw}\{P(W=1|Y=1,X=x)-\Delta_{w}(x)P(Y=1\mid W=1,X=x)\}\mid_{\gamma_x=0}.
\end{split}
\end{equation}
In this case, there is an effect modification due to conditioning of an additional variable, in line with well-known results on non-collapsibility of parameters of logistic regression models, see~\cite{xie2008some}. In addition, we notice that even in this simple case the linearity of $X$ in the marginal model is lost. Furthermore, if $\beta_x$ and $\beta_{xw}$ are both positive (negative), the marginal effect is also positive (negative), thereby recovering the finding in~\cite{cox2003general} on the condition to avoid the effect reversal (i.e., the marginal and conditional effects having opposite signs).

\subsection{Discrete case}\label{subsec:singlewdisc}
\noindent Without loss of generality, we here assume that $X$ is binary. The total effect of $X$ on $Y$ can be derived by taking the first difference of, equivalently,~\eqref{eq:wkfymar1} or~\eqref{eq:wkfymar2}. We here opt for differentiating~\eqref{eq:wkfymar1}. Then
\begin{equation}\label{eq:cochbin}
\tu{TE}(x)=\beta_{x}+ \log \tu{RR}_{\bar W\mid Y, X=0}-\log \tu{RR}_{\bar W\mid Y, X=1}
\end{equation}
which explicitly becomes
\[
\tu{TE}(x)= \beta_x+\log\frac{1+\exp g_{1}(1)}{1+\exp g_{0}(1)} -\log \frac{1+\exp g_{1}(0)}{1+\exp g_{0}(0)}
\]
with $g_y(x)$ as in~\eqref{eq:gyx}. Notice that, in order to make the extension to $X$ discrete straightforward, we maintain the $x$ notation. Obviously, in the case of a binary $X$, $\tu{TE}(x)=\mbox{cpr} (Y,X)$, a constant term corresponding to the cross product ratio of the marginal table for $Y$ and $X$.

\vspace{0.2cm}
\noindent \textbf{Indirect Effect} Following the definition above, we evaluate the total effect assuming  ${\beta_x=\beta_{xw}=0}$, that is
\begin{equation}\label{eq:indbin}
\tu{IE}(x)=\tu{TE}(x)\mid_{\beta_x=\beta_{xw}=0}=\log\frac{1+\exp g^*_{1}(1)}{1+\exp g^*_{0}(1)} -\log \frac{1+\exp g_{1}(0)}{1+\exp g_{0}(0)}
\end{equation}
where $g_{y}^*(x)$ is $g_{y}(x)$ evaluated at $\beta_x=\beta_{xw}=0$. Notice that $g_{y}^*(0)=g_{y}(0)$, see \eqref{eq:gyx}.
In parallel with the continuous case, the indirect effect is null if either $\beta_w$ or $\gamma_x$ are zero. It can be shown with some algebra that it is concordant with the product of the two coefficients.

\vspace{0.2cm}
\noindent \textbf{Direct Effect} 
\noindent Following the definition of the direct effect, we evaluate the direct effect after assuming in the total effect $\beta_w=\beta_{xw}=0$, that is
\begin{equation}\label{eq:dirbin}
\tu{DE}(x)=\tu{TE}(x)\mid_{\beta_w=\beta_{xw}=0}=\beta_x.
\end{equation}

\vspace{0.2cm}
\noindent \textbf{Residual Effect} 
\noindent By definition, the remaining effect is evaluated by difference, such as
\begin{equation}\label{eq:ncebin}
\tu{RES}(x)=\tu{TE}(x)-\tu{IE}(x)-\tu{DE}(x)=\log \frac{1+\exp g_{1}(1)}{1+\exp g_{0}(1)}-\log \frac{1+\exp g^*_{1}(1)}{1+\exp g^*_{0}(1)}
\end{equation}
with $g_{y}^*(x)$ as above. It can easily be seen that $\tu{RES}(x)$ is zero as soon as $\beta_x=\beta_{xw}=0$, leading to $g_{y}^*(1)=g_{y}(1)$, or $\beta_w=\beta_{xw}=0$, leading to $g_{1}(x)=g_{0}(x)$ and $g^*_{1}(x)=g^*_{0}(x)$, see \eqref{eq:gyx}. The latter situation coincides with the condition of collapsibility of odds ratio, see~\cite{xie2008some}, Corollary 3. However, like in the continuous case, there is no clear relationship between the sign of this effect and the logistic regression coefficients.

\vspace{0.2cm}
\noindent \textbf{Some cases of interest}
\noindent Following the definition, we reformulate the total effect of a binary $X$ on a binary $Y$ as in~\eqref{eq:tesum}
and we study in detail the decomposition of the total effect into the indirect \eqref{eq:indbin}, direct \eqref{eq:dirbin} and residual \eqref{eq:ncebin} effects for some cases of interest.

\vspace{0.2cm}
\noindent \textbf{Case \textit{i})} When the recursive logistic models can be depicted as in Fig.~\ref{fig:med}(b), i.e. $X\ind Y\mid W$, it follows from the definition above that 
\[
\tu{TE}(x)\mid_{\beta_x=\beta_{xw}=0}=\tu{IE}(x)
\]
and both $\tu{DE}(x)$ and $\tu{RES}(x)$ are zero.

\vspace{0.2cm}
\noindent \textbf{Case \textit{ii})} When the recursive logistic models can be depicted as in Fig.~\ref{fig:med}(c), i.e. $W\ind Y\mid X$, it follows from the definition above that
\[
\tu{TE}(x)\mid_{\beta_w=\beta_{xw}=0}=\tu{DE}(x),
\]
as all other terms are zero.

\vspace{0.2cm}
\noindent \textbf{Case \textit{iii})} After imposing $\beta_{xw}=0$ the total effect is
\[
\tu{TE}(x)\mid_{\beta_{xw}=0}= \tu{DE}(x) + \tu{IE}(x) + \tu{RES}(x)\mid_{\beta_{xw}=0}
\]
in which
\[
\tu{RES}(x)\mid_{\beta_{xw}=0} = \log \frac{1+\exp g_{1}(1)}{1+\exp g_{0}(1)}\mid_{\beta_{xw}=0}-\log \frac{1+\exp g^*_{1}(1)}{1+\exp g^*_{0}(1)}.
\]
%
%
%
Notice that $g^*_y(x)$ is $g_y(x)$ evaluated at $\beta_x=\beta_{xw}=0$. If we further assume $\gamma_x=0$, after some algebra, it is possible to show that under this condition $|\tu{TE}(x)| \leq |\beta_{x}|$ in line with results obtained by~\cite{neuhaus1993geometric}.

\vspace{0.2cm}
\noindent \textbf{Case \textit{iv})} When the recursive logistic models can be depicted as in Fig.~\ref{fig:med}(d), i.e. $W\ind X$, it follows from the case above
\[
\tu{TE}(x)\mid_{\gamma_x=0}= \tu{DE}(x) + \tu{RES}(x)\mid_{\gamma_x=0}
\]
where
\[
\tu{RES}(x)\mid_{\gamma_x=0} = \log \frac{1+\exp g_{1}(1)}{1+\exp g_{0}(1)}\mid_{\gamma_x=0}-\log \frac{1+\exp g^*_{1}(1)}{1+\exp g^*_{0}(1)}\mid_{\gamma_x=0}.
\]
After some algebra, it is possible to show that this condition is sufficient to avoid effect reversal, as proved by~\cite{cox2003general} in a more general context.

\begin{table}[tb]
\centering
 \resizebox{\textwidth}{!}{  
\begin{tabular}{lccccccccccc}
\toprule
 & \multicolumn{5}{c}{$C=0$} & & \multicolumn{5}{c}{$C=1$}\\
 \cmidrule(l{3pt}r{3pt}){2-6} \cmidrule(l{3pt}r{3pt}){8-12}
 & Est. & SE & \multicolumn{2}{c}{95\% CI} & p-value & & Est. & SE & \multicolumn{2}{c}{95\% CI} & p-value\\ 
 \cmidrule(l{3pt}r{3pt}){2-6} \cmidrule(l{3pt}r{3pt}){8-12}
$\tu{DE}_{\{2,1\}}$   &   1.934 &  0.368&    1.214  &  2.655 &  0.000  &&   1.934   & 0.368 &   1.214  & 2.655 &  0.000 \\
$\tu{IE}_{\{2,1\}}$   &   0.364 &  0.192&   -0.011  &  0.740 &  0.057  &&   0.176   & 0.139 &  -0.096  & 0.449 &  0.205 \\
$\tu{RES}_{\{2,1\}}$  &  -0.476  & 0.197  & -0.862  & -0.089 &  0.016  &&  -0.475 & 0.227 & -0.919 & -0.031 & 0.036 \\
$\tu{TE}_{\{2,1\}}$   &   1.822 &  0.348&    1.141  &  2.506 &  0.000  &&   1.635   & 0.341 &   0.968  & 2.303 &  0.000 \\
\midrule
 & \multicolumn{5}{c}{$C=0$} & & \multicolumn{5}{c}{$C=1$}\\
 \cmidrule(l{3pt}r{3pt}){2-6} \cmidrule(l{3pt}r{3pt}){8-12}
 & Est. & SE & \multicolumn{2}{c}{95\% CI} & p-value & & Est. & SE & \multicolumn{2}{c}{95\% CI} & p-value\\ 
 \cmidrule(l{3pt}r{3pt}){2-6} \cmidrule(l{3pt}r{3pt}){8-12}
$\tu{DE}_{\{3,1\}}$  &  1.133   &0.386  &  0.375&    1.890 & 0.003 &&  1.133  & 0.386  &  0.375 &  1.890 & 0.003\\
$\tu{IE}_{\{3,1\}}$  &  1.475   &0.316  &  0.856&    2.094 & 0.000 &&  0.795  & 0.477  & -0.141 &  1.731 & 0.096\\
$\tu{RES}_{\{3,1\}}$ & -0.846   & 0.300  & -1.435 & -0.257 & 0.005 && -1.057  & 0.567  & -2.168 &  0.054 & 0.062 \\
$\tu{TE}_{\{3,1\}}$  &  1.762   &0.369  &  1.038&    2.486 & 0.000 &&  0.871  & 0.340  &  0.205 &  1.538 & 0.010\\
\bottomrule
\end{tabular}}
\caption{Estimates, standard errors (SEs), 95\% confidence intervals (CIs) and p-values of the effects for the cultural consumption experiment.}
\label{tab:scen2}
\end{table}

With reference to the cultural consumption data, in Table~\mbox{\ref{tab:scen2}} the decomposition of the total effect of moving from level 1 to the other levels of $X$ is reported. Notice that this decomposition is based on the estimated parameters of Table~\mbox{\ref{tab:regr}}, and does not require to estimate the marginal model of $Y$ against $X$ and $C$ only, thereby avoiding the issue of comparing parameters coming from two logistic models with unequal variance. 95$\%$ confidence intervals and p-values are calculated using the approximated standard errors evaluated via the delta method~(\mbox{\Citealt{Oehlert1992}}). 

In the upper part of Table~\mbox{\ref{tab:scen2}}, the decomposition for the $\{2,1\}$ contrast is reported. For both levels of $C$, the total and direct effects are positive and statistically significant, while the indirect effect, also positive, is moderately significant in $C=0$ (p-value 0.057) and non significant for $C=1$ (p-value 0.205).  
As for the $\{3,1\}$ contrast, in the lower part of Table~\mbox{\ref{tab:scen2}}, we see instead that  all the direct, indirect and total effects are positive and statistically significant for both $C=0$ and $C=1$  (though the indirect effect for $C=1$ is only moderately significant). Although, for each contrast, the total, direct and indirect effects are all positive, their interpretation in terms of proportion mediated is not possible given the presence of a negative residual effect; see Section~\mbox{\ref{sec:singlew}}. Such an effect is rather large in magnitude, possibly due to a large interaction coefficient; see Table~\mbox{\ref{tab:regr}}.

In summary, the  direct and total effects of moving from level 1 to level 2 of $X$ are positive and statistically significant in all groups of students, while the indirect effect, also positive, is significant only for students mainly interested in humanities.  When moving from level 1 to level 3 of $X$ all effects are positive and significant.  We believe that this is an important message on how to design incentives to increase museums attendance of high school students, that cannot be easily derived by simply looking at the estimated coefficients in Table~\mbox{\ref{tab:regr}}.

Our results are aligned with the ones in the original studies.~\mbox{\cite{lattarulo2017nudging}} estimated an average causal effect based on the mean difference and the difference in difference methods, marginally with respect to $W$. Instead, in the study of~\mbox{\cite{forastiere2019exploring}}, the authors performed a decomposition of the total effect based on counterfactual entities using the principal stratification method.

\subsection{Effect decomposition on the probability scale}\label{subsec:prob}

So far, we have considered effect decompositions operating on the logistic scale. However, sociologists and econometricians are quite often concerned with effects on the probability scale (also called partial effects; see~\citealt[][Cap. 15]{wooldridge2010econometric}), for which effect decompositions in specific contexts, typically on the additive scale, have also been proposed~\citep{karlson2012comparing,breen2013total}. We here denote with $\eta(x)$ the rhs of \eqref{eq:wkfymar}, that is, 
\begin{equation}
\log\frac{P(Y=1\mid X=x)}{P(Y=0\mid X=x)} = \eta(x)= \log\frac{1+\exp g_{1}(x) }{1+\exp g_{0}(x) } + \beta_{0}+\beta_{x}x.
\end{equation}

For the continuous $X$ case, effects are defined as derivatives w.r. to $x$ of the probability function. The Total Probability Effect ($\textup{TPE}(x)$) is therefore so defined:
\[
\textup{TPE}(x) = \frac{d}{d x}P(Y=1 \mid X=x)=P(Y=1 \mid X=x)\{1-P(Y=1 \mid X=x)\}\textup{TE}(x)
\]
where $\textup{TE}(x)$ corresponds to the total effect on the logistic scale as defined in \eqref{eq:cochranint}. The result follows from the derivative of $\textup{expit}\,\eta(x) = \exp \eta(x)/\{1+\exp \eta(x)\}$ with respect to its argument $\eta(x)$. 

On the other hand, for $X$ binary or discrete, the total effect on the probability scale can be defined by simply taking the difference across levels of $X$ of the marginal probability. For the binary $X$ this becomes:
\[
\textup{TPE}(x) = P(Y=1 \mid X=1)-P(Y=1 \mid X=0)=\mbox{expit} \,\eta(1)- \mbox{expit}\, \eta(0).
\]
In analogy with the approach of Section~\ref{sec:singlew}, the Direct Probability Effect ($\textup{DPE}(x)$) and Indirect Probability Effect ($\textup{IPE}(x)$) are defined by zeroing the corresponding coefficients in $\textup{TPE}(x)$. To obtain an additive decomposition, we also define the Residual Probability Effect ($\textup{RPE}(x)$) by difference as
\[
\mbox{\textup{RPE}}(x) = \mbox{\textup{TPE}}(x) - \mbox{\textup{DPE}}(x) - \mbox{\textup{IPE}}(x).
\]
With particular reference to the continuous case, this amounts to:
\[
\begin{split}
\textup{DPE}(x) &= \textup{TPE}(x) \mid_{\beta_w=\beta_{xw}=0} = \textup{expit}(\beta_0+\beta_xx)\{1-\textup{expit}(\beta_0+\beta_xx)\}\beta_x \\
\textup{IPE}(x) &= \textup{TPE}(x) \mid_{\beta_x=\beta_{xw}=0} = \textup{expit}\{\eta^*(x)\}[1-\textup{expit}\{\eta^*(x)\}]\textup{IE}(x)
\end{split}
\]
where $\textup{IE}(x)$ is as in Eq.~\eqref{eq:iecon} and $\eta^*(x)$ is $\eta(x)$ evaluated at $\beta_x=\beta_{xw}=0$. 
Like the effects on the logistic scale, all these probability effects are local measures, since they depend on the specific value $x$. Global measures can be defined by averaging the aforementioned local quantities. For instance, given a population of $N$ units ($i=1,\dots,N$), one could define the Average Total Probability Effect (ATPE) as
\[
\mbox{\textup{ATPE}}=\frac{1}{N}\sum_{i=1}^N \mbox{\textup{TPE}}(x_i),
\]
with the Average Direct Probability Effect (ADPE) and the Average Indirect Probability Effect (AIPE) defined analogously. However, these average effects should be taken with caution when there is a strong variation across values of $x$.

For the cultural consumption data, the effects on the probability scale are summarized in \mbox{Table~\ref{tab:scen3}}, which has the same structure of \mbox{Table~\ref{tab:scen2}}. As expected, results are in line with the ones on the log odds scale. Notice that averaging the effects may be not appropriate in applications with a strong variation across levels of $x$, as in this case, especially with reference to the indirect effects.

\begin{table}[tb]
	\centering
	\resizebox{\textwidth}{!}{  
		\begin{tabular}{lccccccccccc}
			\toprule
			& \multicolumn{5}{c}{$C=0$} & & \multicolumn{5}{c}{$C=1$}\\
			\cmidrule(l{3pt}r{3pt}){2-6} \cmidrule(l{3pt}r{3pt}){8-12}
			& Est. & SE & \multicolumn{2}{c}{95\% CI} & p-value & & Est. & SE & \multicolumn{2}{c}{95\% CI} & p-value\\ 
			\cmidrule(l{3pt}r{3pt}){2-6} \cmidrule(l{3pt}r{3pt}){8-12}
			$\tu{DPE}_{\{2,1\}}$  &   0.413  &   0.069 &  0.279 &  0.547 &  0.000 &&  0.446 & 0.074 &  0.301 &   0.591 &  0.000\\
			$\tu{IPE}_{\{2,1\}}$  &   0.063  &   0.031 &  0.001 &  0.124 &  0.046 &&  0.035 & 0.028 & -0.020 &   0.090 &  0.215\\
			$\tu{RPE}_{\{2,1\}}$ &  -0.073  &   0.032 & -0.135 & -0.010 &  0.023 && -0.099 & 0.047 & -0.190 &  -0.008 &  0.034\\
			$\tu{TPE}_{\{2,1\}}$  &   0.403  &   0.068 &  0.269 &  0.537 &  0.000 &&  0.382 & 0.072 &  0.240 &   0.523 &  0.000\\
			\midrule
			& \multicolumn{5}{c}{$C=0$} & & \multicolumn{5}{c}{$C=1$}\\
			\cmidrule(l{3pt}r{3pt}){2-6} \cmidrule(l{3pt}r{3pt}){8-12}
			& Est. & SE & \multicolumn{2}{c}{95\% CI} & p-value & & Est. & SE & \multicolumn{2}{c}{95\% CI} & p-value\\ 
			\cmidrule(l{3pt}r{3pt}){2-6} \cmidrule(l{3pt}r{3pt}){8-12}
			$\tu{DPE}_{\{3,1\}}$  &  0.216  &    0.067 &   0.084  &  0.347 &  0.001  &&   0.255 &  0.082 &  0.094 &  0.415 &  0.002\\
			$\tu{IPE}_{\{3,1\}}$  &  0.317  &    0.078 &   0.164  &  0.470 &  0.000  &&   0.176 &  0.119 & -0.058 &  0.410 &  0.141\\
			$\tu{RPE}_{\{3,1\}}$ & -0.144  &    0.074 &  -0.289  &  0.000 &  0.049  &&  -0.236 &  0.154 & -0.539 &  0.067 &  0.127\\
			$\tu{TPE}_{\{3,1\}}$  &  0.388  &    0.091 &   0.210  &  0.566 &  0.000  &&   0.194 &  0.098 &  0.003 &  0.386 &  0.046\\
			\bottomrule
	\end{tabular}}
	\caption{Estimates of the effects on the probability scale for the cultural consumption experiment, with standard errors (SEs), 95\% confidence intervals (CIs) and p-values.}
	\label{tab:scen3}
\end{table}

\section{Simulation study}\label{sec:sims}
In this section, we present results of a simulation study to investigate how well the relative amount of indirect effect is recovered, also in relation with already existing methods. In particular,~\cite{karlson2012comparing} and \cite{breen2013total} derive a decomposition of the total effect that can be applied to the case of a continuous mediator $W$, when the response model for $Y$ can either be a logistic or a probit one with no interaction terms. In this context, the total effect, as measured by the marginal coefficient of $X$ on $Y$, is the sum of the direct and indirect effects, as in linear models. When all effects are positive, it is therefore meaningful to compute the proportion mediated as the ratio between the indirect and total effect; see Eq. (21) in~\cite{breen2013total}. The authors present a method to sidestep the well-known issue of unequal variances, known as KHB method. They also propose to adapt it to the binary mediator case, by postulating a linear probability model for $W$. 

We here postulate a logistic model for $W$ and analyze, through simulations, the behaviour in finite samples of the KHB method and of the proposed method. For $X$ binary,  we compare the KHB method with the ratio between the estimates of the effects $\textup{IE}(x)$ and $\textup{TE}(x)$ in Section~\ref{sec:singlew}, obtained by plugging-in the ML estimates of the parameters in the corresponding expression.
For $X$ continuous, we notice that the KHB measure should be interpreted as the proportion mediated on the probability scale in the same fashion as discussed in Section~\ref{subsec:prob}. For this reason, we proceed as follows. Given a sample of $n$ units, an estimate of the corresponding effect is formed by averaging across units the corresponding entities. As an instance, the estimated Average Total Probability Effect (ATPE) is so formed:
$$
\widehat{\textup{ATPE}}=\frac{1}{n}\sum_{i=1}^n \widehat{\textup{TPE}}(x_i)
$$
where $\widehat{\textup{TPE}}(x_i)$ is the estimated total probability effect of unit $i$, obtained by plugging-in the ML estimates in the corresponding expression. The estimated Average Indirect Probability Effect $(\textup{AIPE})$ is formed accordingly and the ratio between the two entities is then taken. We recall from Section~\ref{sec:singlew} that, due to the presence of the residual effect, this measure should not be interpreted as proportion mediated in the usual way. The above measure is then compared with the KHB measure.

We consider a basic setting with no covariates, where the outcome $Y$ and the mediator $W$ are generated according to Equations~\eqref{eq:lry} and~\eqref{eq:lrw} respectively. Though our method can accommodate for the treatment-mediator interaction $\beta_{xw}$ in the outcome equation, for a fair comparison it is here posed to zero. The remaining parameters are set to: $\beta_0=-2=\gamma_0$ and $\beta_w=2=\gamma_x$, while $\beta_x$ is varying in $\{0.4,0.9,1.8\}$ in order to explore different relative amounts of indirect effect. 

We define three sample sizes, i.e., $n \in \{250,500,1000\}$. In the binary treatment case, the $X$ variate is sampled from a Bernoulli distribution with probability equal to 0.5. In the continuous treatment case, we first generate a large pseudo-population of size $N_{\tu{pop}}=150000$ from a Normal distribution with null mean and variance equal to 2 and then create $X$ by extracting a random sample of size $n$ from it. In this way, the true value of the $\tu{AIPE}/\tu{ATPE}$ ratio is computed on the pseudo-population and does not vary with the sample size. Once $X$ is obtained, for each $n$, $N=2000$ replications of $(W,Y)$ are drawn and estimation is performed. 


\begin{table}[tb]
\centering
\begin{tabularx}{\textwidth}{YccYYYYYYYYY}
\toprule
\multicolumn{3}{r}{$n$} & 250 & 500 & 1000 & 250   & 500  & 1000  & 250 & 500 & 1000 \\
\cmidrule(lr){4-6} \cmidrule(lr){7-9} \cmidrule(lr){10-12}
$\beta_x$ & true value & method & \multicolumn{3}{c}{average} & \multicolumn{3}{c}{variance} & \multicolumn{3}{c}{RMSE} \\
\midrule
 & \multicolumn{11}{l} {$\textup{IE}/\textup{TE}$ ($X$ binary)} \\
\midrule
\multirow{2}{*}{0.4} & \multirow{2}{*}{0.716} & KHB & 0.732 & 0.683 & 0.669 & 0.144 & 0.033 & 0.014 & 0.379 & 0.184 & 0.129 \\
                     	      &  				  & RSD & 0.757 & 0.732 & 0.724 & 0.064 & 0.024 & 0.011 & 0.256 & 0.154 & 0.107 \\
\multirow{2}{*}{0.9} & \multirow{2}{*}{0.532} & KHB & 0.475 & 0.468 & 0.462 & 0.020 & 0.008 & 0.004 & 0.153 & 0.111 & 0.092 \\
                     	      & 				  & RSD & 0.544 & 0.539 & 0.535 & 0.020 & 0.009 & 0.004 & 0.141 & 0.094 & 0.064 \\
\multirow{2}{*}{1.8} & \multirow{2}{*}{0.364} & KHB & 0.301 & 0.300 & 0.297 & 0.004 & 0.002 & 0.001 & 0.092 & 0.079 & 0.074 \\
	                       & 				  & RSD & 0.367 & 0.367 & 0.364 & 0.007 & 0.003 & 0.002 & 0.082 & 0.058 & 0.040 \\
\midrule
 &  \multicolumn{11}{l} {$\textup{AIPE}/\textup{ATPE}$ ($X$ continuous)} \\
\midrule
\multirow{2}{*}{0.4} & \multirow{2}{*}{0.590} & KHB & 0.531 & 0.521 & 0.513 & 0.028 & 0.013 & 0.006 & 0.178 & 0.133 & 0.108 \\
                     	     & 			                   & RSD & 0.561 & 0.589 & 0.580 & 0.010 & 0.004 & 0.002 & 0.103 & 0.064 & 0.042 \\
\multirow{2}{*}{0.9} & \multirow{2}{*}{0.437} & KHB & 0.316 & 0.317 & 0.310 & 0.008 & 0.004 & 0.002 & 0.149 & 0.135 & 0.134 \\
			      & 			           & RSD & 0.411 & 0.458 & 0.440 & 0.002 & 0.002 & 0.001 & 0.055 & 0.046 & 0.026 \\
\multirow{2}{*}{1.8} & \multirow{2}{*}{0.351} & KHB & 0.180 & 0.178 & 0.180 & 0.002 & 0.001 & 0.001 & 0.178 & 0.177 & 0.173 \\
                     	      & 		                   & RSD & 0.328 & 0.343 & 0.352 & 0.001 & 0.001 & 0.000 & 0.042 & 0.026 & 0.017 \\        
\bottomrule
\end{tabularx}
\caption{True value and simulation average, variance and Root Mean Squared Error (RMSE) for the KHB method and the proposed method (RSD).}\label{tab:sims}
\end{table}

Table~\ref{tab:sims} summarizes the simulation results. Notice that for $X$ binary, the dependence of $\textup{IE}$ and $\textup{TE}$ on $x$ is removed. As expected, the proposed estimators always approach the true values as the sample size grows, with smaller Root Mean Squared Error (RMSE) in all scenarios considered. Conversely, the KHB estimator is biased, with RMSE increasing with the value of $\beta_x$. Furthermore, if for the binary case the bias seems reasonable, it becomes consistent in the continuous case. 


\section{Extension to multiple binary mediators}\label{sec:multiplew}

\begin{figure}[tb]
\centering{
\begin{tikzpicture}[scale=0.4,auto,->,>=stealth',shorten >=1pt,node distance=2cm] 
\node[littlenode] (Wk) {$W_k$};
\node[littlenode,inner sep=0] (Wk-1) [right of=Wk]{$W_{k-1}$};
\node (Wj) [right of=Wk-1]{$\dots$};
\node[littlenode] (W1) [right of=Wj] {$W_1$};
\node[littlenode] (X) [below of=Wk] {$X$};
\node[littlenode] (Y) [below of=W1] {$Y$};
\draw[->] (X) --node {} (Y) ;
\draw[->] (X) --node {} (Wk) ;
\draw[->] (X) --node {} (Wk-1) ;
\draw[->] (X) --node {} (Wj) ;
\draw[->] (X) --node {} (W1) ;

\draw[->] (Wk) --node {} (Y) ;
\draw[->] (Wk-1) --node {} (Y) ;
\draw[->] (Wj) --node {} (Y) ;
\draw[->] (W1) --node {} (Y) ;

\draw[->] (Wk)  --node {} (Wk-1) ;
\draw[->] (Wk) edge[bend left] node {} (Wj) ;
\draw[->] (Wk) edge[bend left] node {} (W1) ;

\draw[->] (Wk-1) --node {} (Wj) ;
\draw[->] (Wk-1) edge[bend left] node {} (W1) ;

\draw[->] (Wj) edge[bend left] node {} (W1) ;
\draw[->] (Wj) --node {} (W1) ;


\end{tikzpicture}
\caption{DAG with $k$ mediators.\label{fig:medk}}
}
\end{figure}
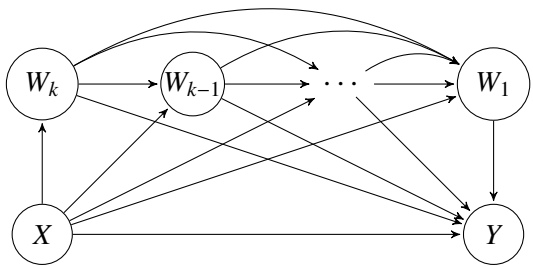

The proposed definitions of direct, indirect and residual effects, together with their parametric formulations, extend nicely to the situation where multiple binary mediators are present. Suppose there are $k$ mediators and that a full ordering among the variables $(Y, W_1, W_{k-1}, \ldots, W_k,X)$ is available such that each variable is a potential response variable for the subsequent ones. The system can be represented via a DAG, see  Figure~\ref{fig:medk}. We assume that each response model is a hierarchical logistic model. In the following, if we impose the regression coefficient of one covariate to be zero, all higher order interaction terms involving this covariate are implicitly imposed to zero.
For brevity, we denote with $W_{>j}$, the set of all $W_r$ such that $r>j$. 
The coefficients of $X$ and of $W_j$ and of their interactions in the logistic regression of $Y$ are denoted with $\beta$, in a self-explaining fashion.

In analogy with~\eqref{eq:wkfymar1}, the logistic model for $Y$ given $X$, obtained after marginalization upon the $k$ mediators, is
\begin{equation}
\begin{aligned}
\log \frac{P(Y=1\mid X=x)}{P(Y=0\mid X=x)}&=\beta_0 + \beta_{x}x - \sum_{j=1}^{k} \log \tu{RR}_{\bar{W_j}\mid Y,X=x,W_{>j}=0} 
\end{aligned}\label{eq:general}
\end{equation}
\noindent where 
$$
\tu{RR}_{\bar W_j\mid Y,X=x,W_{>j}=w_{>j}}=\frac{1+\exp g_{0}^{(w_{<j})}(x,w_{>j})}{1+\exp g_{1}^{(w_{<j})}(x,w_{>j}) }.
$$
In Appendix~\ref{app:mult} the relevant expressions are given.

\vspace{0.2cm}
In line with what done for the simple case, we here offer a generalization of the definitions for the total, direct, indirect and residual effects under the situation of $k$ mediators.

\vspace{0.2cm}
\noindent {\bf Total Effect} 
\noindent Let $\tu{TE}(x)$ be the total effect of $X$ on $Y$ on the log odds scale, after marginalization on $k$ binary mediators. For $X$ continuous and differentiable, the total effect is defined as the derivative of~\eqref{eq:general} with respect to $x$. It follows that
\begin{equation}\label{eq:TEmultiplecon}
\begin{aligned}
\tu{TE}(x)&=\beta_x+\sum_{j=1}^{k} \frac{\exp g^{(w_{<j})}_1 (x,w_{>j}=0)}{1+\exp g^{(w_{<j})}_1 (x,w_{>j}=0)} \frac{d}{d x}g^{(w_{<j})}_1 (x,w_{>j}=0) \\
& - \sum_{j=1}^{k} \frac{\exp g^{(w_{<j})}_0 (x,w_{>j}=0)}{1+\exp g^{(w_{<j})}_0 (x,w_{>j}=0)} \frac{d}{d x} g^{(w_{<j})}_0 (x,w_{>j}=0).
\end{aligned}
\end{equation}

For $X$ discrete, the total effect is defined as the difference between~\eqref{eq:general} evaluated at two different levels of $X$. Without loss of generality, we here assume $X$ binary, and take the difference for $x=1$ and $x=0$. Then
\begin{equation}\label{eq:TEmultiple}
\begin{aligned}
\tu{TE}(x)&=\beta_x+ \sum_{j=1}^{k} \log \frac{\tu{RR}_{\bar{W_j}\mid Y,X=0,W_{>j}=0}}{\tu{RR}_{\bar{W_j}\mid Y,X=1,W_{>j}=0}}.
\end{aligned}
\end{equation}
Equation~\eqref{eq:TEmultiple} can also be written as follows
\begin{equation}\notag
\begin{aligned}
\tu{TE}(x) &=\beta_x+ \sum_{j=1}^{k} \log \frac{1+\exp g^{(w_{<j})}_1 (1,w_{>j}=0)}{1+\exp g^{(w_{<j})}_0 (1,w_{>j}=0)} - \sum_{j=1}^{k} \log \frac{1+\exp g^{(w_{<j})}_1 (0,w_{>j}=0)}{1+\exp g^{(w_{<j})}_0 (0,w_{>j}=0)}.
\end{aligned}
\end{equation}

\vspace{0.2cm}
\noindent {\bf Direct Effect}
\noindent Let $\tu{DE}(x)$ be the direct effect of $X$ on $Y$ on the log odds scale. Let $\bar \beta_{W}$ be the set of all $\beta$ regression coefficients of each mediator in the model for $Y$, including also the interaction terms both between mediators and between mediators and $X$. The direct effect is evaluated in $\tu{TE}(x)$ after imposing $\bar \beta_{W}=0$, thereby $Y \ind W_1, \ldots, W_k \mid X$, i.e.

\begin{equation}\label{eq:DEmultiple}
\begin{aligned}
\tu{DE}(x)&=\tu{TE}(x)\mid_{\bar \beta_{W}=0}=\beta_x.
\end{aligned}
\end{equation}

\vspace{0.2cm}
\noindent {\bf Global Indirect Effect} The global indirect effect $\tu{GIE}(x)$ can be defined in analogy with the previous definitions, as the total effect evaluated when the direct effect of $X$, $\beta_x$, is zero. Since we only deal with hierarchical models, this implies that all interaction terms between $X$ and the mediators are also zero.  Therefore,
\begin{equation}\label{eq:TIEmultiple}
\tu{GIE}(x)=\tu{TE}(x)\mid _{\beta_x=0}.
\end{equation}

\vspace{0.2cm}
\noindent {\bf Residual Effect} 
\noindent Let $\tu{RES}(x)$ be the residual effect evaluated by difference, i.e.,
\begin{equation}\label{eq:NCEmultiple}
\tu{RES}(x)=\tu{TE}(x)-\tu{GIE}(x)-\tu{DE}(x).
\end{equation}

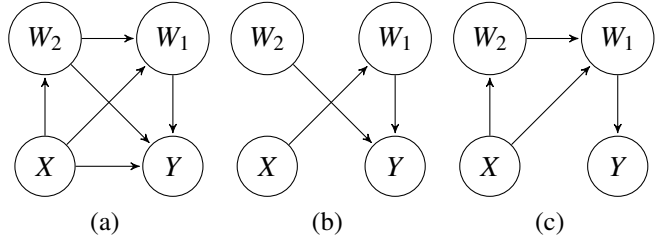
\begin{figure}[tb]
\centering{
\subfloat[][]{\label{fig:dag2_1}
\begin{tikzpicture}[scale=0.35,auto,->,>=stealth',shorten >=.9pt,node distance=1.7cm] 
\node[littlenode] (W2) {$W_2$};
\node[littlenode] (W1) [right of=W2] {$W_1$};
\node[littlenode] (X) [below of=W2] {$X$};
\node[littlenode] (Y) [below of=W1] {$Y$};
\draw[->] (W2) --node {} (Y) ; \draw[->] (X) --node {} (W1); \draw[->] (X) --node {} (Y); \draw[->] (X) --node {} (W2);  \draw[->] (W2) --node {} (W1); \draw[->] (W1) --node {} (Y) ;
\end{tikzpicture}}\,
\subfloat[][]{\label{fig:dag2_2}
\begin{tikzpicture}[scale=0.35,auto,->,>=stealth',shorten >=.9pt,node distance=1.7cm] 
\node[littlenode] (W2) {$W_2$};
\node[littlenode] (W1) [right of=W2] {$W_1$};
\node[littlenode] (X) [below of=W2] {$X$};
\node[littlenode] (Y) [below of=W1] {$Y$};
\draw[->] (W2) --node {} (Y) ; \draw[->] (X) --node {} (W1);   \draw[->] (W1) --node {} (Y) ;
\end{tikzpicture}}\,
\subfloat[][]{\label{fig:dag2_3}
\begin{tikzpicture}[scale=0.35,auto,->,>=stealth',shorten >=.9pt,node distance=1.7cm] 
\node[littlenode] (W2) {$W_2$};
\node[littlenode] (W1) [right of=W2] {$W_1$};
\node[littlenode] (X) [below of=W2] {$X$};
\node[littlenode] (Y) [below of=W1] {$Y$};
 \draw[->] (X) --node {} (W1);  \draw[->] (X) --node {} (W2);  \draw[->] (W2) --node {} (W1); \draw[->] (W1) --node {} (Y) ;
\end{tikzpicture}}\,
\caption{DAGs with $k=2$ mediators when (a) no conditional independences hold (b) $X \ind Y \mid \{W_1,W_2\}$,  $W_2 \ind W_1 \mid X$ and $X \ind W_2$ (c) $Y \ind \{X,W_2\} \mid W_1$.\label{fig:med2}}
}
\end{figure}

\noindent From previous derivations, non-zero residual effects are induced by graphs having more than one arrow pointing to $Y$. Therefore, we can state that the residual effect is zero whenever one of the two following graphical conditions holds: (i) there is no direct path from $X$ to $Y$ or (ii) there is the direct path from $X$ to $Y$ and no other arrow is pointing to $Y$. As an instance, the model corresponding to the DAG in Fig.~\ref{fig:med2}(a) has a non-zero $\tu{RES}(x)$ as there is a direct arrow form $X$ to $Y$ and other two arrows are pointing to $Y$, while models corresponding to DAGs as in Figs.~\ref{fig:med2}(b) and \ref{fig:med2}(c) are such that $\tu{TE}(x)=\tu{GIE}(x)$ and $\tu{RES}(x)=0$.

In a setting with multiple mediators, one is also interested in a path-specific indirect effect, i.e. the effect that is due to some mediators only, and is null whenever one arrow along the pathway is deleted. Notice that, in this setting, also other research questions are of interest, such as the path-specific indirect effects when some mediators are marginalized over. They are addressed in Section~\ref{sec:other}.

\vspace{0.2cm}
\noindent {\bf Path-Specific Indirect Effect} Let $A$ be one of the $2^{k-1}$ ordered subsets of $(W_1,W_2,\ldots,W_k)$ containing at least one element of $W$. Let $i_A$ be the ordered set of indices $j$ such that $W_j \in A$. The path-specific indirect effect $\tu{PSIE}_A(x)$ is obtained from the total effect after imposing that:
\begin{itemize}
\item $\beta_x=0$;
\item $\beta_{w_j}=0$ with $j=\{1,2,\ldots,k\}, j \neq \mbox{min}\{i_A\}$ the smallest index in $i_A$;
\item $\gamma_{r,j}=0$  with $W_r \in A$, $j>r, j \neq \ell_r$, where $\ell_r$ is the index following $r$ in $i_A$; 
\item $\gamma_{r,x}=0$ with $W_r \in A$, $r \neq \mbox{max}\{i_A\}$ the largest index in $i_A$.
\end{itemize}
In this way, each path-specific indirect effect contains only the parameters pertaining to the path (including the intercepts). It then follows that the indirect effect $\tu{PSIE}_A(x)$ is null whenever one of the following conditions holds:
\begin{itemize}
\item $\beta_{w_j}=0$ with $j= \mbox{min}\{i_A\}$;
\item $\gamma_{r,\ell_r}=0$ with $W_r \in A$;
\item $\gamma_{r,x}$ with $r = \mbox{max}\{i_A\}$.
\end{itemize}
Notice that each of the conditions above implies deleting one arrow in the DAG corresponding to the model of interest. 

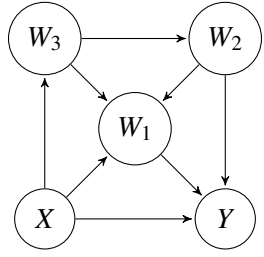
\begin{figure}[tb]
\centering{
\begin{tikzpicture}[scale=0.4,auto,->,>=stealth',shorten >=1pt,node distance=2cm] 
\node[littlenode] (W3) at (0,6) {$W_3$};
\node[littlenode] (W2) at (6,6) {$W_2$};
\node[littlenode] (W1) at (3,3) {$W_1$};
\node[littlenode] (X) at (0,0) {$X$};
\node[littlenode] (Y) at (6,0) {$Y$};
\draw[->] (X) --node {} (Y) ;
\draw[->] (X) --node {} (W1) ;
\draw[->] (X) --node {} (W3) ;
\draw[->] (W3) --node {} (W1) ;
\draw[->] (W3) --node {} (W2) ;
\draw[->] (W2) --node {} (W1) ;
\draw[->] (W2) --node {} (Y) ;
\draw[->] (W1)  --node {} (Y) ;
\end{tikzpicture}
}
\caption{DAG with $W_1$ acting as a collider node in the path $X\rightarrow W_3 \rightarrow W_1 \leftarrow W_2 \rightarrow Y$.  \label{fig:med7}}
\end{figure}

As an instance, let $k=6$, $i_A=\{2,3,5\}$. The path-specific indirect effect is obtained from $\tu{TE}(x)$ after imposing that:
\begin{itemize}
\item $\beta_x=\beta_{xw_1}=\ldots=\beta_{xw_6}=0$;
\item $\beta_{w_1}=\beta_{w_3}=\beta_{w_4}=\beta_{w_5}=\beta_{w_6}=0$ ;
\item $\gamma_{2,4}=\gamma_{2,5}=\gamma_{2,6}=\gamma_{3,4}=\gamma_{3,6}=\gamma_{5,6}=0$;
\item $\gamma_{2,x}=	\gamma_{3,x}=0$
\end{itemize}
\noindent where $\gamma_{j,x}$ and $\gamma_{j,i}$ are, in order, the coefficients of $X$ and $W_i$ in the equation of $W_j$ against its parent nodes in the corresponding DAG. 
The above definition allows for only one path from $X$ to $Y$, which is 
$X \rightarrow  W_5\rightarrow W_{3}\rightarrow W_2 \rightarrow Y$. It is null whenever $\gamma_{5,x}$ or $\gamma_{3,5}$ or $\gamma_{2,3}$ or $\beta_{w_2}$ are zero.

Notice that the definition of path-specific indirect effects allows only for direction preserving paths, i.e. paths with all arrows pointing to the same direction. As a matter of fact, only ordered subsets of $W$ are allowed to form $A$. This choice is justified by the fact that these are the only subsets with a non-zero path-specific indirect effect. To clarify the issue, see the graph in Fig.~\ref{fig:med7}. The path $X\rightarrow W_3 \rightarrow W_1 \leftarrow W_2 \rightarrow Y$ is not admitted as it gives rise to $(W_3,W_1,W_2)$, not an ordered subset of $W$. However, as $W_1$ is a collider node the path between $X$ and $Y$ is blocked by 
$(W_3,W_1,W_2)$ and the corresponding path-specific effect is zero, see~\Citet*[Chs. 1 and 3]{PearlCausalityBook2009}.

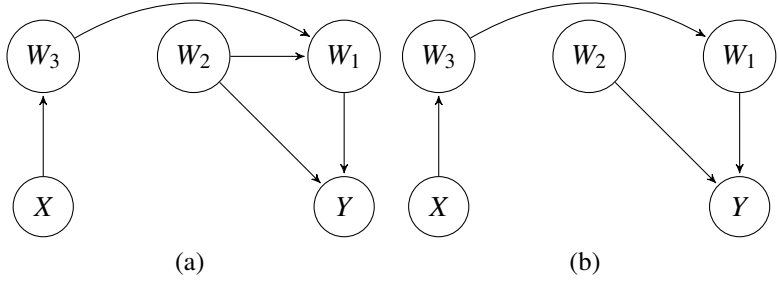
\begin{figure}[tb]
\centering{
\subfloat[][]{\label{fig:dag3_1}
\begin{tikzpicture}[scale=0.35,auto,->,>=stealth',shorten >=1pt,node distance=2cm] 
\node[littlenode] (W3) {$W_3$};
\node[littlenode] (W2) [right of=W3]{$W_2$};
\node[littlenode] (W1) [right of=W2] {$W_1$};
\node[littlenode] (X) [below of=W3] {$X$};
\node[littlenode] (Y) [below of=W1] {$Y$};
\draw[->] (W2) --node {} (W1); 
\draw[->] (W2) --node {} (Y) ; 
\draw[->] (W3) edge[bend left] node {} (W1); 
\draw[->] (X) --node {} (W3); 
\draw[->] (W1) --node {} (Y) ; 
\end{tikzpicture}}\,
\subfloat[][]{\label{fig:dag3_2}
\begin{tikzpicture}[scale=0.35,auto,->,>=stealth',shorten >=1pt,node distance=2cm] 
\node[littlenode] (W3) {$W_3$};
\node[littlenode] (W2) [right of=W3]{$W_2$};
\node[littlenode] (W1) [right of=W2] {$W_1$};
\node[littlenode] (X) [below of=W3] {$X$};
\node[littlenode] (Y) [below of=W1] {$Y$};
\draw[->] (W2) --node {} (Y) ; 
\draw[->] (W3) edge[bend left] node {} (W1); 
\draw[->] (X) --node {} (W3); 
\draw[->] (W1) --node {} (Y) ; 
\end{tikzpicture}}\,
\caption{DAG with $k=3$ mediators (a) $Y \ind \{X,W_3\}  \mid \{W_1,W_2\}$, $ W_1 \ind X \mid \{W_2,W_3\}$, $W_2 \ind X$ (b) $Y \ind \{X,W_3\}  \mid \{W_1,W_2\}$, $ W_1 \ind \{X,W_2\} \mid W_3$, $W_2 \ind X$.\label{fig:med3}}
}
\end{figure}
It is also important to notice that, in parallel to  the single-mediator case, $\tu{DE}(x)$ coincides with the effect of $X$ on $Y$ keeping $W_1=W_2=\ldots=W_k=0$. However, $\tu{PSIE}_A(x)$ does not in general coincide with the indirect effect after keeping the mediators not in $A$ equal to zero. To see this, notice that in Fig.~\ref{fig:med7} the path-specific indirect effect for $A=(W_3,W_2)$ is evaluated after imposing that $\beta_x=\beta_{w_1}=0$. This effect does not coincide with the one obtained after conditioning on $W_1=0$, see~\cite{elwert2014endogenous}.

Notice that the sum of all the path-specific indirect effects in general is not equal to the global indirect effect. This is true even when there is just one path from $X$ to $Y$. This is due to the different ways to deal with the effects induced in non-collapsible subgraphs. These are subgraphs involving three random variables,  $(W_i,W_j,W_r)$,  or $(W_i,W_j,Y)$, $i >j>r$, such that there are two arrows pointing to the inner node, i.e. $W_r$ or $Y$.  In this case $W_j$ acts as a mediator between $W_i$ and $W_r$, or $Y$, and there is a non-zero residual effect; see Section~\mbox{\ref{sec:singlew}}. Specifically, the global indirect effect includes all residual effects, whereas path-specific indirect effects do not. 

As an example, consider the models with DAG as in Figures~\ref{fig:med3}(a) and~\ref{fig:med3}(b). In both DAGs there is just one indirect path leading from $X$ to $Y$, i.e. $ X \rightarrow W_3 \rightarrow W_1 \rightarrow Y$, with $A=(W_3,W_1)$. Both models have $\tu{GIE}(x) \neq \tu{PSIE}_A(x)$. The model corresponding to Figures~\mbox{\ref{fig:med3}} (a) has two non-collapsible subgraphs, namely the ones induced by $(W_3,W_2,W_1)$ and $(W_2,W_1, Y)$, while the model corresponding to Figure~\mbox{\ref{fig:med3}} (b) has one non-collapsible subgraph, namely the one induced by $(W_2,W_1, Y) $. Notice the different meaning of the parameters attached to the arrow $W_3\rightarrow W_1$ in the two effects: in the $\tu{GIE}(x)$ it is the total effect of $W_3$ on $W_1$, while in the path-specific $\tu{PSIE}_A(x)$ it is the direct effect.



\section{Other path-specific indirect effects of interest} \label{sec:other}

Suppose now that the research question involves path-specific effects in the model obtained after marginalization over some mediators while others are kept in the model.  First of all, the parameters of the marginal model of interest should be obtained  and then the path-specific indirect effects can be evaluated. Two different situations may arise. The first one involves marginalization over an inner mediator, and therefore~\eqref{eq:general} can be used in a straightforward manner. The second one involves marginalization over one intermediate/outer node and more technicalities are necessary. We here present an instance of both situations. 

Suppose that the research question involves investigation of the path-specific indirect effects in the model obtained after marginalization over $W_1$ of a model corresponding to the DAG in Figure~\ref{fig:med8}(a). In Figure~\ref{fig:med8}(b), the DAG corresponding to the marginal model of interest is presented, with the red arrows corresponding to parameters that change due to the marginalization over $W_1$. The expressions for these parameters is reported in Appendix~\ref{app:psie}. The only non-zero path-specific indirect effects are for $A=\{W_2\}$ and $A=\{W_3\}$. They can be evaluated by making use of the parameters of the marginal model.  Notice that, as expected, the $\tu{GIE}(x)$ in the marginal model is equal to the $\tu{GIE}(x)$ in the original model.

\begin{figure}[tb]
\centering{
\subfloat[][]{\label{fig:dag8_2}
\begin{tikzpicture}[scale=0.35,auto,->,>=stealth',shorten >=1pt,node distance=1.8cm] 
\node[littlenode] (W3) {$W_3$};
\node[littlenode] (W2) [right of=W3]{$W_2$};
\node[littlenode, negate] (W1) [right of=W2] {$W_1$};
\node[littlenode] (X) [below of=W3] {$X$};
\node[littlenode] (Y) [below of=W1] {$Y$};
\draw[->] (W2) --node {} (W1); 
\draw[->] (W2) --node {} (Y) ; 
\draw[->] (W3) edge[bend left] node {} (W1); 
\draw[->] (X) --node {} (W3); 
\draw[->] (W1) --node {} (Y) ; 
\draw[->] (X) --node {} (Y); 
\draw[->] (X) --node {} (W2); 
\end{tikzpicture}}\,
\subfloat[][]{\label{fig:dag9_2}
\begin{tikzpicture}[scale=0.20,auto,->,>=stealth',shorten >=1pt,node distance=1.8cm] 
\node[littlenode] (W3) {$W_3$};
\node[littlenode] (W2) [right of=W3]{$W_2$};
\node[littlenode] (X) [below of=W3] {$X$};
\node[littlenode] (Y) [below of=W1] {$Y$};
\draw[->] (X) --node {} (W3); 
\draw[->] (X) --node {} (W2);
\color{red} \draw[->] (X) --node {} (Y); 
\color{red} \draw[->] (W3) --node {} (Y); 
\color{red} \draw[->] (W2) --node {} (Y) ; 
\end{tikzpicture}}
\caption{(a) Marginalization over the inner mediator $W_1$ and (b) quantifying the parameters (in red the parameters that change). \label{fig:med8}}
}
\end{figure}
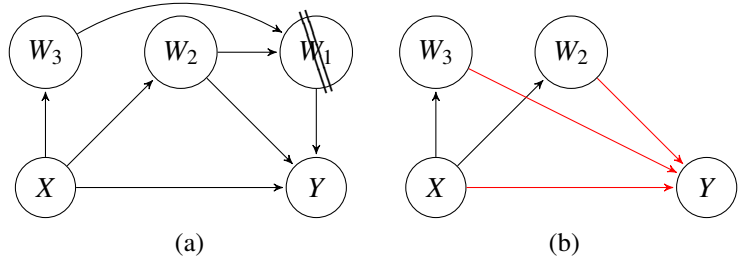

\begin{figure}[tb]
\centering{
\subfloat[][]{\label{fig:dag10_1}
\begin{tikzpicture}[scale=0.35,auto,->,>=stealth',shorten >=.9pt,node distance=1.7cm] 
\node[littlenode, negate] (W2) {$W_2$};
\node[littlenode] (W1) [right of=W2] {$W_1$};
\node[littlenode] (X) [below of=W2] {$X$};
\node[littlenode] (Y) [below of=W1] {$Y$};
\draw[->] (W2) --node {} (Y) ; \draw[->] (X) --node {} (W1); \draw[->] (X) --node {} (Y); \draw[->] (X) --node {} (W2);  \draw[->] (W2) --node {} (W1); \draw[->] (W1) --node {} (Y) ;
\end{tikzpicture}}\,
\subfloat[][]{\label{fig:dag10_2}
\begin{tikzpicture}[scale=0.35,auto,->,>=stealth',shorten >=.9pt,node distance=1.7cm] 
\node[littlenode] (W1) {$W_1$};
\node[littlenode] (X) [below left of=W1] {$X$};
\node[littlenode] (Y) [below right of=W1] {$Y$};
\color{red}\draw[->] (X) --node {} (W1);   \draw[->] (W1) --node {} (Y) ; \draw[->] (X) --node {} (Y);
\end{tikzpicture}}\,
\caption{(a) Marginalization over the outer mediator $W_2$ and (b) quantifying the parameters (in red the parameters that change).\label{fig:med10}}
}
\end{figure}
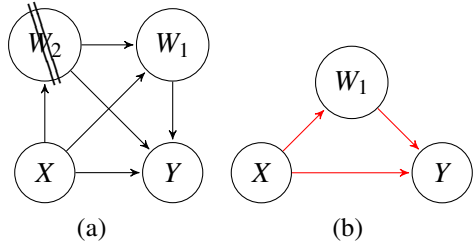

Quantification of effects in models obtained after marginalization over intermediate or outer nodes involves repeated use of the derivations here presented.  We here detail the steps to be followed for the case with $k=2$ mediators. Generalizations to more complex models can be derived after repeatedly applying the procedure here proposed. 

Suppose that we wish to evaluate the indirect effect in the model with $W_1$ as unique mediator, i.e. the model obtained after marginalization over $W_2$, see Figure~\ref{fig:med10}. This implies deriving the parametric formulation of 
\begin{equation}\label{eq.E4}
\begin{aligned}
\log \frac{P(Y=1\mid X=x,W_1=w_1)}{P(Y=0\mid X=x,W_1=w_1)}&=\log\frac{P(Y=1\mid X=x,W_1=w_1,W_2=0)}{P(Y=0\mid X=x,W_1=w_1,W_2=0)}\\
&+\log\frac{P(W_2=0\mid Y=0,X=x,W_1=w_1)}{P(W_2=0\mid Y=1,X=x,W_1=w_1)}.
\end{aligned}
\end{equation}
\noindent in which the second term of the right hand side of the equation is to be determined. From repeated use of the derivations in Appendix~\ref{app:cov}, we have
\begin{equation}\label{eq.E5}
\begin{aligned}
\log \frac{P(Y=1\mid X=x,W_1=w_1)}{P(Y=0\mid X=x,W_1=w_1)}&=\beta_0+\beta_x x + \beta_{w_1}w_1 + \beta_{xw_1}xw_1+\log\frac{1+\exp h_{1,w_1}(x)}{1+\exp h_{0,w_1}(x)}.
\end{aligned}
\end{equation}
with the expression of $h_{y,w_1}$ in Appendix~\ref{app:last}. The values of the marginal parameters are straightforward; see e.g. Appendix~\ref{app:psie}.


\section{Causal interpretation of total, direct and indirect effects}\label{sec:causal}

In the counterfactual framework, many approaches exist to mediation analysis, and a review is in~\cite{huber2019review}. In a single mediator context,~\cite{VdWVan2010} define the counterfactual notion of direct and indirect effects when the outcome is binary, thereby focusing on the log odds scale. Within a regression analysis context with a continuous mediator, the authors present an approximated parametric formulation of the effects that holds under the rare outcome assumption of $Y$.~\cite{valeri2013mediation} address the same problem when also the mediator is binary, again modeled under the rare outcome assumption. It is therefore worth to explore the links existing between the effects introduced here and these causal effects defined in a formal counterfactual framework. Since the latter are contrasts expressed, possibly after a logarithmic transformation, by a difference, this parallel holds for $X$ binary. Notice that, differently from the above cited approaches, we here present a decomposition based on the exact formulation of the effects on the log odds scale. 

Under the assumption that the recursive system of equation is structural in the sense of~\Citet*[Ch. 7]{PearlCausalityBook2009}, one can give the total effect and some of its components a causal interpretation. To say that the recursive system of equations is structural  implies that the DAG is a causal graph that satisfies a set of axioms, namely composition, effectiveness and reversibility, see also~\Citet*{steen2018mediation}. 

With a single binary mediator, a parallelism between the structural definition of a DAG and the Sequential Ignorability Assumption of~\Citet*{imai2010identification} exists, see~\cite{pearl2012mediation} and~\Citet*{shpitser2011complete}. Under the assumption of no unmeasured counfounder of the treatment-outcome relationship, possibly after conditioning on a set of pre-treatment covariates $C$, the total effect of $X$ on $Y$ here presented corresponds to the \textit{total causal effect} as defined by~\cite{VdWVan2010}. Similarly, assuming that there are no unobserved confounders of the treatment-outcome relationship, possibly after conditioning on a set of pre-treatment covariates $C$,  and no unmeasured confounders of the mediator-outcome relationship, after conditioning on the treatment $X$ and possibly some pre-treatment covariates $C$, the direct effect can be seen as the \textit{controlled direct effect}~(\Citealt{VdWVan2010}) after an external intervention to fix $W=0$ is performed (see Section 2.2 case $ii)$). Less obvious are the parallelisms in terms of natural effects of~\cite{VdWVan2010}. It is possible to show that the \textit{pure natural indirect effect} can be seen as the total effect after assuming $\beta_{x}=\beta_{xw}=0$ (i.e. the indirect effect; see Section 2.2 case $i)$) and that the \textit{pure natural direct effect} corresponds to the total effect after assuming $\gamma_x=0$, i.e. $X \ind W$ (see Section 2.2 case $iv)$). More details are in~\cite{doretti2018exact}.

When multiple causally ordered mediators are present, several possible effects are of interest, see~\cite{daniel2015causal,steen2017flexible}. However, in this case the definition of natural direct and indirect effects is more cumbersome, and sometimes effects of interest are non-parametrically non identifiable, see e.g. the situation described in~\Citet*{avin2005identifiability}. Again, if no unobserved confounders exist, some parallelisms continue to hold. As an instance, the direct effect can be seen as the controlled direct effect of $X$ on $Y$ after an external intervention to fix the mediator $W_1=W_2=\ldots =W_k=0$ is performed, see also~\cite{vanderweele2014mediation}. Our approach allows further to appreciate the total and controlled direct effect of $X$ on $Y$ also when some mediators are marginalized over, while others are kept in the system. We believe that this is an important research question in many applied studies. 


\section{Conclusions}\label{sec:concl}

Logistic regression is by far the most used model for a binary response. Further than in a mediation context, the models here proposed may arise in longitudinal studies with a binary outcome measured at different occasions.

With reference to a single mediator, we have proposed a novel decomposition of the total effect into direct and indirect effects that is more appropriate for the nonlinear case and that, under certain conditions, reduces to the classical definition in the linear case. Additionally, this decomposition overcomes the issue of unequal variance when fitting two nested models. As illustrative example, we have re-analyzed data based on an encouragement program to stimulate students' attitude to visit museums. We have shown how the decomposition of the total effect could avoid erroneous conclusions on the direct and indirect effects and provides additional information that cannot be found by just looking at the results of the separate regressions analysis. Although the total, direct and indirect effects are all positive, a substantial residual effect, possibly due to  a large interaction coefficient, hinders the interpretation in terms of proportion mediated.

Additional important results concern the extension of the definitions to the multiple mediator context. Repeated use of the decomposition of the total effect allows to address complex issues like quantifying the total, direct and indirect effects when a subset of mediators are marginalized over. Links to the causal effects have also been established.


\appendix
\section{Appendix: a general formulation}\label{app:cov}

The definitions of total, direct, indirect effects can be extended to include a set of covariates $C=(C_1,C_2,\ldots, C_p)$. Let $c=(c_1,c_2,\ldots,c_p)$ denote a possible value of $C$. Let $\rhs(Y\mid W=w,X=x,C=c)$ denote the 
right hand side of the logit model for $Y$ against $X$ and $C$. Notice that $\rhs(Y\mid W=1,X=0,C=0)$ is the part $\rhs(Y\mid W=w,X=x,C=c)$ that contains $w$ terms only.  Let $\rhs(W\mid X=x,C=c)$ denote the right hand side of the logit model for $W$ against $X$ and $C$. 

A straightforward extension of the derivations presented in Section 1 leads to the following:
\begin{equation}\label{eq:all}
\begin{split}
 \log \frac{P(Y=1\mid X=x,C=c)}{P(Y=0\mid X=x,C=c)}= \rhs(Y\mid W=1,X=x,C=c) -\log \tu{RR}_{W\mid Y,X=x,C=c}\\
 = \rhs(Y\mid W=0,X=x,C=c) -\log \tu{RR}_{\bar{W}\mid Y,X=x,C=c}
\end{split}
\end{equation}
with
\[
\begin{split}
\log \tu{RR}_{W\mid Y,X=x,C=c} &= \log \frac{\exp g_{1}(x, c)}{1+\exp g_{1}(x, c)} - \log\frac{\exp g_{0}(x,c)}{1+\exp g_{0}(x,c)}\\
\log \tu{RR}_{\bar{W}\mid Y,X=x,C=c} &= \log\frac{1+\exp g_{0}(x,c) }{1+\exp g_{1}(x, c)},
\end{split}
\]
in which
\begin{equation}\label{eq:gyxall}
\begin{split}
g_y(x,c)& =y\{\rhs(Y\mid W=1,X=x,C=c)-\rhs(Y\mid W=0,X=x,C=c) \} \\
&+\log\frac{1+\exp \{\rhs(Y\mid W=0,X=x,C=c)\}}{1+\exp\{\rhs(Y\mid W=1,X=x,C=c)\}} + \rhs(W\mid X=x,C=c).
\end{split}
\end{equation}

\noindent 
The total effect can be obtained from Equations~\eqref{eq:all} - \eqref{eq:gyxall}, after taking the derivative with respect to $X$ (for $X$ continuous) or taking difference among levels of $X$ (for $X$ discrete).


\section{Appendix: concordance between \mbox{$\Delta_w^*(x)$} and \mbox{$\beta_w$}}\label{app:delta}

To prove that $\Delta^*_w(x)$ and $\beta_w$ are concordant, we first prove the concordance between $\Delta_y(x)$ and $\Delta_w(x)$. For $x$ fixed, one could specify a $2 \times 2$ table with entries $p_{wy}=P(W=w,Y=y \mid X=x)$ so that $\Delta_y(x)=(p_{00}p_{11} - p_{01}p_{10})/\{(p_{11}+p_{10})(p_{00}+p_{01})\}$ and $\Delta_w(x)=(p_{00}p_{11} - p_{01}p_{10})/\{(p_{11}+p_{01})(p_{00}+p_{10})\}$. Thus 
\[
\Delta_w(x) = \Delta_y(x)\frac{(p_{11}+p_{10})(p_{00}+p_{01})}{(p_{01}+p_{11})(p_{00}+p_{10})},
\]
with the second factor on the right-hand side always positive, which proves concordance. As this result holds in general, also $\Delta_w^*(x)$ and $\Delta_y^*(x)$ are concordant. The result follows after noticing that $\textup{expit}(\cdot)$ is a monotone function, therefore $\beta_w$ shares the sign with $\Delta_y^*(x)=\textup{expit}(\beta_0+\beta_w) - \textup{expit}(\beta_0)$.


\section{Appendix: multiple mediators}\label{app:mult}

We here generalize the previous derivations to a situation with multiple mediators. We denote with $g_y^{(w_{<j})}(x, w_{>j})$ the function~ \eqref{eq:gyxall}  evaluated after marginalizing over the $W_{<j}$ mediators and considering $(W_{j+1},\ldots,W_k)$ as covariates. The expression of $g_y^{(w_{<j})}(x, w_{>j})$ can be derived recursively, after noting that $g_y(x,w_{>1})$ is~\eqref{eq:gyxall} after imposing $W=W_1$ and $C=W_{>1}$. 
We marginalize iteratively over the inner mediator, that is starting from the marginalization over $W_1$, we have
\begin{equation}
\begin{aligned}
\rhs (Y\mid X=x, W_2=w_2, \ldots, W_k=w_k)&=\rhs (Y\mid X=x, W_1=0, W_2=w_2, \ldots, W_k=w_k)\\
&-\log \tu{RR}_{\bar{W_1}\mid Y,X=x,W_{>1}=w_{>1}}
\end{aligned}
\end{equation}
and then over $W_2$, we have
\begin{equation}
\begin{aligned}
\rhs (Y\mid X=x, W_3=w_3, \ldots, W_k=w_k)&=\rhs (Y\mid X=x, W_1=0, W_2=0,W_3=w_3, \ldots, W_k=w_k)\\
&-\log \tu{RR}_{\bar{W_1}\mid Y,W_2=0, X=x,W_{>2}=w_{>2}}\\
& - \log \tu{RR}_{\bar{W_2}\mid Y, X=x, W_{>2}=w_{>2}}
\end{aligned}
\end{equation}
\noindent up to marginalization over $W_k$, that is
\begin{equation}
\begin{aligned}
\rhs (Y\mid X=x)=\rhs (Y\mid X=x, W_1=0, \ldots W_k=0)-\sum_{j=1}^k \log \tu{RR}_{\bar{W_j}\mid Y,X=x,W_{>j}=0}
\end{aligned}
\end{equation}
\noindent where, at each step, 
$$
\log \tu{RR}_{\bar W_j\mid Y,X=x,W_{>j}}=\log\frac{1+\exp g_0^{(w_{<j})}(x,w_{>j})}{1+\exp g_1^{(w_{<j})}(x,w_{>j})}
$$
and
\begin{equation}\label{eq:gyxallmarg}
\begin{split}
g_y^{(w_{<j})}(x,w_{>j})& =y\{\rhs(Y\mid W_j=1,X=x,W_{>j}=w_{>j})-\rhs(Y\mid W_j=0,X=x,W_{>j}=w_{>j}) \} \\
&+\log\frac{1+\exp \{\rhs(Y\mid W_j=0,X=x,W_{>j}=w_{>j})\}}{1+\exp\{\rhs(Y\mid W_j=1,X=x,W_{>j}=w_{>j})\}} + \rhs(W_j\mid X=x,W_{>j}=w_{>j}).
\end{split}
\end{equation}

\section{Appendix: marginalization over $W_1$ with $k=3$}\label{app:psie}

Let us consider the model with $k=3$ and $X$ binary. No conditional independences are assumed. We here report the parametric forms for the parameters changed after the marginalization over $W_1$. Let us consider four logistic regression models, respectively, for the outcome $Y$, and for the three mediators $W_1$, $W_2$, $W_3$, with interaction terms up to the second order. 

We denote with $\gamma_{j,0}$, $\gamma_{j,x}$, $\gamma_{j,i}$, $\gamma_{j,xi}$, $\gamma_{j,hi}$, in order, the intercept, the coefficient of $X$, the coefficient of $W_i$, the interaction term of $X$ and $W_i$ and the interaction term of $W_h$ and $W_i$ in the regression of $W_j$ against $W_r$, $r>j$, and $X$. We also denote with $\beta_0^{(w_i)}$, $\beta_x^{(w_i)}$, etc., in order, the intercept, the main effect of $X$, etc., of the logistic model for $Y$ against the covariates after marginalization on $W_i$ and $\beta_0^{(w_i,w_{>j})}$, $\beta_x^{(w_i,w_{>j})}$, etc., in order, the intercept, the main effect of $X$, etc., of the logistic model for $Y$ against the covariates after marginalization on $W_i$ and $W_{>j}$. 
Notice that marginalization over the inner mediator induces changes in the parameters of the outcome equation only. From Eq.~\eqref{eq:all}, considering $W_2$ and $W_3$ as covariates and marginalizing over $W_1$ we obtain the first marginal model for $Y$ against $X, W_1, W_2, W_3$ over $W_1$ such as

\begin{equation}\notag
\begin{aligned}
\log\frac{P(Y=1\mid X=x, W_2=w_2, W_3=w_3)}{P(Y=0\mid X=x,W_2=w_2, W_3=w_3)} &=\beta_0^{(w_1)} + \beta_x^{(w_1)} x + \sum_{j=2}^k \beta_{w_j}^{(w_1)}w_j  \\
&+ \sum_{j=2}^k \beta_{xw_j}^{(w_1)}xw_j+  \beta_{xw_2 w_3}^{(w_1)}xw_2w_3\\
&= \beta_0 + \beta_x x + \beta_{w_2}w_2 + \beta_{xw_2}xw_2\\
&  - \log \tu{RR}_{\bar{W_1} \mid Y,X=x, W_2=w_2, W_3=w_3}
\end{aligned}
\end{equation}
\noindent where 
$$
\log \tu{RR}_{\bar{W_1} \mid Y,X=x, W_2=w_2, W_3=w_3}=\log \frac{1+\exp g_{0}(x,w_2,w_3)}{1+\exp g_{1}(x, w_2,w_3)}
$$
with $g_y(x,w_2,w_3)$ obtained from \eqref{eq:gyxall} after imposing $W=W_1$ and $C=(W_2,W_3)$. It then follows that 
\begin{equation}\notag
\begin{aligned}
\beta_{0}^{(w_1)}=\beta_{0} - \log \tu{RR}_{\bar{W_1} \mid Y,X=0, W_2=0, W_3=0}.
\end{aligned}
\end{equation}
The main effects are
\begin{equation}\notag
\begin{aligned}
\beta_x^{(w_1)}&=\log  \cpr ({Y,X}\mid W_2=0, W_3=0)\\
&=\beta_x+ \log \tu{RR}_{\bar{W_1}\mid Y, X=0, W_2=0, W_3=0} -  \log \tu{RR}_{\bar{W_1}\mid Y, X=1, W_2=0, W_3=0}
\end{aligned}
\end{equation}
\begin{equation}\notag
\begin{aligned}
\beta_{w_2}^{(w_1)}&=\log  \cpr ({Y,W_2}\mid X=0, W_3=0)\\
&=\beta_{w_2}+
\log \tu{RR}_{\bar{W_1}\mid Y, X=0, W_2=0, W_3=0} -  \log \tu{RR}_{\bar{W_1}\mid Y, X=0, W_2=1, W_3=0}
\end{aligned}
\end{equation}
\begin{equation}\notag
\begin{aligned}
\beta_{w_3}^{(w_1)}&=\log  \cpr ({Y,W_3}\mid X=0, W_2=0)\\
&=\beta_{w_3} + \log \tu{RR}_{\bar{W_1}\mid Y, X=0, W_2=0, W_3=0} -  \log \tu{RR}_{\bar{W_1}\mid Y, X=0, W_2=0, W_3=1}.
\end{aligned}
\end{equation}
The second order interactions are
\begin{equation}\notag
\begin{aligned}
\beta_{xw_2}^{(w_1)}&=\log  \cpr ({Y,X}\mid W_2=1, W_3=0)-\log  \cpr ({Y,X}\mid W_2=0, W_3=0)\\
&=\beta_{xw_2}+ \log \frac{ \tu{RR}_{\bar{W_1}\mid Y, X=0, W_2=1, W_3=0}}{\tu{RR}_{\bar{W_1}\mid Y, X=1, W_2=1, W_3=0}} + \log \frac{ \tu{RR}_{\bar{W_1}\mid Y, X=1, W_2=0, W_3=0}}{\tu{RR}_{\bar{W_1}\mid Y, X=0, W_2=0, W_3=0}},\\
\end{aligned}
\end{equation}
\begin{equation}\notag
\begin{aligned}
\beta_{xw_3}^{(w_1)}&=\log  \cpr ({Y,X}\mid  W_2=0, W_3=1)-\log  \cpr ({Y,X}\mid  W_2=0,W_3=0)\\
&= \beta_{xw_3} + \log \frac{ \tu{RR}_{\bar{W_1}\mid Y, X=0,W_2=0, W_3=1 }}{\tu{RR}_{\bar{W_1}\mid Y, X=1,  W_2=0,W_3=1}} + \log \frac{ \tu{RR}_{\bar{W_1}\mid Y, X=1, W_2=0,W_3=0 }}{\tu{RR}_{\bar{W_1}\mid Y, X=0, W_2=0,W_3=0}},\\
\end{aligned}
\end{equation}
\begin{equation}\notag
\begin{aligned}
\beta_{w_2w_3}^{(w_1)}&=\log  \cpr ({Y,W_2}\mid  X=0, W_3=1)-\log  \cpr ({Y,W_2}\mid  X=0,W_3=0)\\
&= \beta_{w_2w_3} +  \log \frac{ \tu{RR}_{\bar{W_1}\mid Y, X=0,W_2=0, W_3=1 }}{\tu{RR}_{\bar{W_1}\mid Y, X=0,  W_2=1,W_3=1}} + \log \frac{ \tu{RR}_{\bar{W_1}\mid Y, X=0, W_2=1,W_3=0 }}{\tu{RR}_{\bar{W_1}\mid Y, X=0, W_2=0,W_3=0}}.\\
\end{aligned}
\end{equation}
The third order interaction is
\begin{equation}\notag
\begin{aligned}
\beta_{xw_2w_3}^{(w_1)}&=\left[\log  \cpr ({Y,X}\mid  W_2=1, W_3=1)-\log  \cpr ({Y,X}\mid  W_2=1,W_3=0)\right]\\
&-\left[ \log  \cpr ({Y,X}\mid  W_2=0, W_3=1)-\log  \cpr ({Y,X}\mid  W_2=0,W_3=0) \right]\\
&= \log \frac{ \tu{RR}_{\bar{W_1}\mid Y, X=0, W_2=1,W_3=1 }}{\tu{RR}_{\bar{W_1}\mid Y, X=1, W_2=1,W_3=1}}-\log \frac{ \tu{RR}_{\bar{W_1}\mid Y, X=0, W_2=1,W_3=0 }}{\tu{RR}_{\bar{W_1}\mid Y, X=1, W_2=1,W_3=0}}\\
&-\log \frac{ \tu{RR}_{\bar{W_1}\mid Y, X=0, W_2=0,W_3=1 }}{\tu{RR}_{\bar{W_1}\mid Y, X=1, W_2=0,W_3=1}}+\log \frac{ \tu{RR}_{\bar{W_1}\mid Y, X=0, W_2=0,W_3=0 }}{\tu{RR}_{\bar{W_1}\mid Y, X=1, W_2=0,W_3=0}}.
\end{aligned}
\end{equation}

\noindent Theses quantities are generic for models without any restrictions in terms of conditional independence assumptions. Therefore, for the specific example depicted in Fig.~\ref{fig:med8}, the values of the marginalized parameters must take into account the conditional independences of the corresponding models. 

\noindent Notice that even if we did not include any type of interaction in the full model, after the first marginalization, the interactions of second and third order appear.



\section{Appendix: marginalization over $W_2$ with $k=2$}\label{app:last}

\noindent From repeated use of the first principles of probability it follows that
\begin{equation}\label{eq.E1}
\begin{aligned}
\log \frac{P(W_2=1\mid Y=y,X=x,W_1=w_1)}{P(W_2=0\mid Y=y,X=x,W_1=w_1)}&=\log\frac{P(Y=y\mid X=x,W_1=w_1,W_2=1)}{P(Y=y\mid X=x,W_1=w_1,W_2=0)}\\
&+\log\frac{P(W_2=1\mid X=x,W_1=w_1)}{P(W_2=0\mid X=x,W_1=w_1)}
\end{aligned}
\end{equation}
and
\begin{equation}\label{eq.E2}
\log\frac{P(W_2=1\mid X=x,W_1=w_1)}{P(W_2=0\mid X=x,W_1=w_1)}=\log\frac{P(W_1=w_1\mid X=x,W_2=1)}{P(W_1=w_1\mid X=x,W_2=0)} + \log \frac{P(W_2=1\mid X=x)}{P(W_2=0 \mid X=x)}.
\end{equation}
Denoting with $h_{y,w_1}(x)$ the left hand side of~\eqref{eq.E1} we have
\begin{equation}\label{eq.E3}
\begin{aligned}
h_{y,w_1}(x)=&\log\frac{P(Y=y\mid X=x,W_1=w_1,W_2=1)}{P(Y=y\mid X=x,W_1=w_1,W_2=0)} + \log\frac{P(W_1=w_1\mid X=x,W_2=1)}{P(W_1=w_1\mid X=x,W_2=0)}\\
&+ \log \frac{P(W_2=1\mid X=x)}{P(W_2=0 \mid X=x)}
\end{aligned}
\end{equation}
where all the quantities in~\eqref{eq.E3} are known from the assumed models. In particular, following the notation used in Appendix~\ref{app:cov}, we obtain that
\begin{equation}
\begin{aligned}
\log\frac{P(Y=y\mid X=x,W_1=w_1,W_2=1)}{P(Y=y\mid X=x,W_1=w_1,W_2=0)}&=y\{\rhs(Y\mid W_2=1,X=x,W_1=w_1)\\
&-\rhs(Y\mid W_2=0,X=x,W_1=w_1) \} \\
&+\log\frac{1+\exp \{\rhs(Y\mid W_2=0,X=x,W_1=w_1)\}}{1+\exp\{\rhs(Y\mid W_2=1,X=x,W_1=w_1)\}},
\end{aligned}
\end{equation}
\begin{equation}
\begin{aligned}
\log\frac{P(W_1=w_1\mid X=x,W_2=1)}{P(W_1=w_1\mid X=x,W_2=0)}=&w_1\{\rhs(W_1\mid W_2=1,X=x)-\rhs(W_1\mid W_2=0,X=x) \} \\
&+\log\frac{1+\exp \{\rhs(W_1\mid W_2=0,X=x)\}}{1+\exp\{\rhs(W_1\mid W_2=1,X=x)\}}
\end{aligned}
\end{equation}
and
\begin{equation}
\log \frac{P(W_2=1\mid X=x)}{P(W_2=0 \mid X=x)}=\rhs(W_2\mid X=x)=\gamma_{2,0}+\gamma_{2,x}x.
\end{equation}



\begin{thebibliography}{}

\bibitem[\protect\citeauthoryear{Alwin and Hauser}{Alwin and
  Hauser}{1975}]{alwin1975decomposition}
Alwin, D.~F. and R.~M. Hauser (1975).
\newblock The decomposition of effects in path analysis.
\newblock {\em American Sociological Review\/}~{\em 40\/}(1), 37--47.

\bibitem[\protect\citeauthoryear{Avin, Shpitser, and Pearl}{Avin
  et~al.}{2005}]{avin2005identifiability}
Avin, C., I.~Shpitser, and J.~Pearl (2005).
\newblock Identifiability of path-specific effects.
\newblock In {\em Proceedings of International Joint Conference on Artificial
  Intelligence, Edinburgh, Schotland}, pp.\  357--363.

\bibitem[\protect\citeauthoryear{Baron and Kenny}{Baron and
  Kenny}{1986}]{baron1986moderator}
Baron, R.~M. and D.~A. Kenny (1986).
\newblock The moderator--mediator variable distinction in social psychological
  research: Conceptual, strategic, and statistical considerations.
\newblock {\em Journal of Personality and Social Psychology\/}~{\em 51\/}(6),
  1173.

\bibitem[\protect\citeauthoryear{Bollen}{Bollen}{1987}]{bollen1987total}
Bollen, K.~A. (1987).
\newblock Total, direct, and indirect effects in structural equation models.
\newblock {\em Sociological Methodology\/}~{\em 17}, 37--69.

\bibitem[\protect\citeauthoryear{Breen, Karlson, and Holm}{Breen
  et~al.}{2013}]{breen2013total}
Breen, R., K.~B. Karlson, and A.~Holm (2013).
\newblock Total, direct, and indirect effects in logit and probit models.
\newblock {\em Sociological Methods \& Research\/}~{\em 42\/}(2), 164--191.

\bibitem[\protect\citeauthoryear{Breen, Karlson, and Holm}{Breen
  et~al.}{2018}]{breen2018note}
Breen, R., K.~B. Karlson, and A.~Holm (2018).
\newblock A note on a reformulation of the {KHB} method.
\newblock {\em Sociological Methods \& Research\/}, Available at
  \url{https://doi.org/10.1177/0049124118789717}.

\bibitem[\protect\citeauthoryear{Cochran}{Cochran}{1938}]{Cochran1938}
Cochran, W.~G. (1938).
\newblock The omission or addition of an independent variate in multiple linear
  regression.
\newblock {\em Supplement to J. R. Statist. Soc.\/}~{\em 5\/}(2), 171--176.

\bibitem[\protect\citeauthoryear{Cox and Wermuth}{Cox and
  Wermuth}{2003}]{cox2003general}
Cox, D. and N.~Wermuth (2003).
\newblock A general condition for avoiding effect reversal after
  marginalization.
\newblock {\em Journal of the Royal Statistical Society: Series B (Statistical
  Methodology)\/}~{\em 65\/}(4), 937--941.

\bibitem[\protect\citeauthoryear{Daniel, De~Stavola, Cousens, and
  Vansteelandt}{Daniel et~al.}{2015}]{daniel2015causal}
Daniel, R., B.~De~Stavola, S.~Cousens, and S.~Vansteelandt (2015).
\newblock Causal mediation analysis with multiple mediators.
\newblock {\em Biometrics\/}~{\em 71\/}(1), 1--14.

\bibitem[\protect\citeauthoryear{Doretti, Raggi, and Stanghellini}{Doretti
  et~al.}{2021}]{doretti2018exact}
Doretti, M., M.~Raggi, and E.~Stanghellini (2021).
\newblock Exact parametric causal mediation analysis for a binary outcome with
  a binary mediator.
\newblock {\em Statistical Methods \& Applications (forthcoming)\/.}

\bibitem[\protect\citeauthoryear{Elwert}{Elwert}{2013}]{elwert2013graphical}
Elwert, F. (2013).
\newblock Graphical causal models.
\newblock In {\em Handbook of causal analysis for social research}, pp.\
  245--273. Springer.

\bibitem[\protect\citeauthoryear{Elwert and Winship}{Elwert and
  Winship}{2014}]{elwert2014endogenous}
Elwert, F. and C.~Winship (2014).
\newblock Endogenous selection bias: The problem of conditioning on a collider
  variable.
\newblock {\em Annual Review of Sociology\/}~{\em 40}, 31--53.

\bibitem[\protect\citeauthoryear{Forastiere, Lattarulo, Mariani, Mealli, and
  Razzolini}{Forastiere et~al.}{2019}]{forastiere2019exploring}
Forastiere, L., P.~Lattarulo, M.~Mariani, F.~Mealli, and L.~Razzolini (2019).
\newblock Exploring encouragement, treatment, and spillover effects using
  principal stratification, with application to a field experiment on teens'
  museum attendance.
\newblock {\em Journal of Business \& Economic Statistics\/}, 1--15.

\bibitem[\protect\citeauthoryear{Huber}{Huber}{2019}]{huber2019review}
Huber, M. (2019).
\newblock A review of causal mediation analysis for assessing direct and
  indirect treatment effects.
\newblock Technical report, Universit{\'e} de Fribourg.

\bibitem[\protect\citeauthoryear{Imai, Keele, and Yamamoto}{Imai
  et~al.}{2010}]{imai2010identification}
Imai, K., L.~Keele, and T.~Yamamoto (2010).
\newblock Identification, inference and sensitivity analysis for causal
  mediation effects.
\newblock {\em Statistical Science\/}~{\em 25\/}(1), 51--71.

\bibitem[\protect\citeauthoryear{Karlson, Holm, and Breen}{Karlson
  et~al.}{2012}]{karlson2012comparing}
Karlson, K.~B., A.~Holm, and R.~Breen (2012).
\newblock Comparing regression coefficients between same-sample nested models
  using logit and probit: A new method.
\newblock {\em Sociological Methodology\/}~{\em 42\/}(1), 286--313.

\bibitem[\protect\citeauthoryear{Lattarulo, Mariani, and Razzolini}{Lattarulo
  et~al.}{2017}]{lattarulo2017nudging}
Lattarulo, P., M.~Mariani, and L.~Razzolini (2017).
\newblock Nudging museums attendance: a field experiment with high school
  teens.
\newblock {\em Journal of Cultural Economics\/}~{\em 41\/}(3), 259--277.

\bibitem[\protect\citeauthoryear{Lauritzen}{Lauritzen}{1996}]{lauritzen1996graphical}
Lauritzen, S.~L. (1996).
\newblock {\em Graphical models}, Volume~17.
\newblock Clarendon Press.

\bibitem[\protect\citeauthoryear{Lin, Psaty, Bruce, and Kronmal}{Lin et~al.}{1998}]{Linetal1998}
Lin, D.~Y., B.~M. Psaty, M. Bruce, and R.~A. Kronmal (1998).
\newblock Assessing the sensitivity of regression results to unmeasured confounders in observational studies.
\newblock {\em Biometrics\/}~{\em 54\/}(3),
  948--963.

\bibitem[\protect\citeauthoryear{Lupparelli}{Lupparelli}{2019}]{lupparelli2019conditional}
Lupparelli, M. (2019).
\newblock Conditional and marginal relative risk parameters for a class of
  recursive regression graph models.
\newblock {\em Statistical Methods in Medical Research\/}~{\em 28\/}(10--11),
  3466--3486.

\bibitem[\protect\citeauthoryear{MacKinnon, Lockwood, Brown, Wang, and
  Hoffman}{MacKinnon et~al.}{2007}]{mackinnon2007intermediate}
MacKinnon, D.~P., C.~M. Lockwood, C.~H. Brown, W.~Wang, and J.~M. Hoffman
  (2007).
\newblock The intermediate endpoint effect in logistic and probit regression.
\newblock {\em Clinical Trials\/}~{\em 4\/}(5), 499--513.

\bibitem[\protect\citeauthoryear{Neuhaus and Jewell}{Neuhaus and
  Jewell}{1993}]{neuhaus1993geometric}
Neuhaus, J.~M. and N.~P. Jewell (1993).
\newblock A geometric approach to assess bias due to omitted covariates in
  generalized linear models.
\newblock {\em Biometrika\/}~{\em 80\/}(4), 807--815.

\bibitem[\protect\citeauthoryear{Oehlert}{Oehlert}{1992}]{Oehlert1992}
Oehlert, G.~W. (1992).
\newblock A note on the delta method.
\newblock {\em The American Statistician\/}~{\em 46\/}(1), 27--29.

\bibitem[\protect\citeauthoryear{Pearl}{Pearl}{2001}]{Pearl2001}
Pearl, J. (2001).
\newblock {Direct and Indirect Effects}.
\newblock In {\em Proceedings of the 17\textsuperscript{th} International
  Conference on Uncertainty in Artificial Intelligence}, UAI'01, San Francisco,
  pp.\  411--420. Morgan Kaufmann Publishers Inc.

\bibitem[\protect\citeauthoryear{Pearl}{Pearl}{2009}]{PearlCausalityBook2009}
Pearl, J. (2009).
\newblock {\em Causality: Models, Reasoning, and Inference\/}
  (2\textsuperscript{nd} ed.).
\newblock New York, NY, USA: Cambridge University Press.

\bibitem[\protect\citeauthoryear{Pearl}{Pearl}{2012}]{pearl2012mediation}
Pearl, J. (2012).
\newblock {\em The mediation formula: A guide to the assessment of causal
  pathways in nonlinear models}.
\newblock Wiley Online Library.

\bibitem[\protect\citeauthoryear{Shpitser and VanderWeele}{Shpitser and
  VanderWeele}{2011}]{shpitser2011complete}
Shpitser, I. and T.~J. VanderWeele (2011).
\newblock A complete graphical criterion for the adjustment formula in
  mediation analysis.
\newblock {\em The International Journal of Biostatistics\/}~{\em 7\/}(1).

\bibitem[\protect\citeauthoryear{Stanghellini and Doretti}{Stanghellini and
  Doretti}{2019}]{stanghellini2019marginal}
Stanghellini, E. and M.~Doretti (2019).
\newblock On marginal and conditional parameters in logistic regression models.
\newblock {\em Biometrika\/}~{\em 106\/}(3), 732--739.

\bibitem[\protect\citeauthoryear{Steen, Loeys, Moerkerke, and
  Vansteelandt}{Steen et~al.}{2017}]{steen2017flexible}
Steen, J., T.~Loeys, B.~Moerkerke, and S.~Vansteelandt (2017).
\newblock Flexible mediation analysis with multiple mediators.
\newblock {\em American Journal of Epidemiology\/}~{\em 186\/}(2), 184--193.

\bibitem[\protect\citeauthoryear{Steen and Vansteelandt}{Steen and
  Vansteelandt}{2018}]{steen2018mediation}
Steen, J. and S.~Vansteelandt (2018).
\newblock Mediation analysis.
\newblock In {\em Handbook of Graphical Models}, pp.\  407--439. CRC Press.

\bibitem[\protect\citeauthoryear{Valeri and VanderWeele}{Valeri and
  VanderWeele}{2013}]{valeri2013mediation}
Valeri, L. and T.~J. VanderWeele (2013).
\newblock Mediation analysis allowing for exposure--mediator interactions and
  causal interpretation: theoretical assumptions and implementation with {SAS}
  and {SPSS} macros.
\newblock {\em Psychological Methods\/}~{\em 18\/}(2), 137--150.

\bibitem[\protect\citeauthoryear{VanderWeele and Vansteelandt}{VanderWeele and
  Vansteelandt}{2014}]{vanderweele2014mediation}
VanderWeele, T. and S.~Vansteelandt (2014).
\newblock Mediation analysis with multiple mediators.
\newblock {\em Epidemiologic Methods\/}~{\em 2\/}(1), 95--115.

\bibitem[\protect\citeauthoryear{VanderWeele and Vansteelandt}{VanderWeele and
  Vansteelandt}{2010}]{VdWVan2010}
VanderWeele, T.~J. and S.~Vansteelandt (2010).
\newblock Odds ratios for mediation analysis for a dichotomous outcome.
\newblock {\em American Journal of Epidemiology\/}~{\em 172\/}(12), 1339--1348.

\bibitem[\protect\citeauthoryear{Winship and Mare}{Winship and
  Mare}{1983}]{winship1983structural}
Winship, C. and R.~D. Mare (1983).
\newblock Structural equations and path analysis for discrete data.
\newblock {\em American Journal of Sociology\/}~{\em 89\/}(1), 54--110.

\bibitem[\protect\citeauthoryear{Wooldridge}{Wooldridge}{2010}]{wooldridge2010econometric}
Wooldridge, J.~M. (2010).
\newblock {\em Econometric Analysis of Cross Section and Panel Data\/}
  (2\textsuperscript{nd} ed.).
\newblock Cambridge, Massachusetts, USA: MIT Press.

\bibitem[\protect\citeauthoryear{Xie, Ma, and Geng}{Xie
  et~al.}{2008}]{xie2008some}
Xie, X., Z.~Ma, and Z.~Geng (2008).
\newblock Some association measures and their collapsibility.
\newblock {\em Statistica Sinica\/}~{\em 18\/}(3), 1165--1183.

\end{thebibliography}

\end{document}